\documentclass[journal]{IEEEtran}
\ifCLASSINFOpdf
  \usepackage[pdftex]{graphicx}
\else
\fi
\usepackage{amsmath}
\ifCLASSOPTIONcompsoc
 \usepackage[caption=false,font=normalsize,labelfont=sf,textfont=sf]{subfig}
\else
 \usepackage[caption=false,font=footnotesize]{subfig}
\fi

\usepackage[dvipsnames]{xcolor}
\usepackage[binary-units]{siunitx}
\usepackage{glossaries}     
\usepackage{booktabs}       
\usepackage{threeparttable} 
\usepackage{multicol}       
\usepackage{multirow}       
\usepackage{url}
\usepackage{adjustbox}
\usepackage[hidelinks]{hyperref}
\usepackage{orcidlink}
\usepackage{cleveref}       

\usepackage{tikz}
\usepackage[dvipsnames]{xcolor}
\usetikzlibrary{arrows,arrows.meta}

\definecolor{cieee0}{HTML}{00629b}
\definecolor{cieee1}{RGB}{255,199, 44}
\definecolor{cieee2}{RGB}{232,119, 34}
\definecolor{cieee3}{RGB}{186, 12, 47}

\DeclareSIUnit{\gateequi}{GE}
\DeclareSIUnit{\op}{Op}
\DeclareSIUnit{\b}{b}
\DeclareSIUnit{\param}{}

\newcommand{\mtnote}[1]{\textsuperscript{\TPTtagStyle{#1}}} 
\newcommand{\x}{$\times$}

\newcommand\marsellus{\textsc{Marsellus}}
\newcommand\rbe{{\gls{rbe}}}

\newcommand{\riscv}{RISC-V}

\newcommand\kin{$\mathrm{K_{in}}$}

\newacronym{sw}{SW}{software}
\newacronym{hw}{HW}{hardware}
\newacronym{soa}{SoA}{state-of-the-art}
\newacronym{hpc}{HPC}{High Performance Computing}
\newacronym{iot}{IoT}{Internet of Things}
\newacronym{vlsi}{VLSI}{Very-Large-Scale Integration}

\newacronym{ar}{AR}{Augmented Reality}
\newacronym{vr}{VR}{Virtual Reality}

\newacronym{tpu}{TPU}{tensor processing units}
\newacronym{tdp}{TDP}{Thermal Design Power}

\newacronym{ml}{ML}{Machine Learning}
\newacronym{ai}{AI}{Artificial Intelligence}
\newacronym{dl}{DL}{Deep Learning}
\newacronym{nn}{NN}{Neural Network}
\newacronym{qnn}{QNN}{Quantized Neural Network}
\newacronym{bnn}{BNN}{Binary Neural Network}
\newacronym{tnn}{TNN}{Ternary Neural Network}
\newacronym{ann}{ANN}{Artificial Neural Network}
\newacronym{dnn}{DNN}{Deep Neural Network}
\newacronym{fcnn}{FCNN}{Fully Connected Neural Networks}
\newacronym{fcl}{FCL}{Fully Connected Layer}
\newacronym{cnn}{CNN}{Convolutional Neural Network}
\newacronym{rnn}{RNN}{Recurrent Neural Network}
\newacronym{gru}{GRU}{Gated Recurrent Unit}
\newacronym{lstm}{LSTM}{Long Short-Term Memory}
\newacronym{mlp}{MLP}{Multi-Layer Perceptron}
\newacronym{rl}{RL}{Reinforcement Learning}

\newacronym{relu}{ReLU}{Rectified Linear Unit}

\newacronym{ic}{IC}{integrated circuit}
\newacronym[plural=SoCs,firstplural=Systems-on-a-Chip (SoCs)]{soc}{SoC}{System-on-a-Chip}
\newacronym{hesoc}{heSoC}{heterogeneous System-on-a-Chip}
\newacronym{fpga}{FPGA}{field-programmable gate array}
\newacronym{asic}{ASIC}{application-specific integrated circuit}
\newacronym{asip}{ASIP}{Application-Specific Instruction-Set Processor}
\newacronym{cpu}{CPU}{Central Processing Unit}
\newacronym{gpu}{GPU}{Graphics Processing Unit}
\newacronym{gpgpu}{GPGPU}{General-Purpose \acrlong{gpu}}
\newacronym{wse}{WSE}{wafer-scale engine}
\newacronym{pcb}{PCB}{Printed Circuit Board}
\newacronym{lsb}{LSB}{least significant bit}
\newacronym{msb}{MSB}{most significant bit}

\newacronym{sram}{SRAM}{Static Random Access Memory}
\newacronym{dram}{DRAM}{Dynamic Random Access Memory}
\newacronym{noc}{NoC}{Network-on-a-Chip}
\newacronym{isa}{ISA}{Instruction Set Architecture}
\newacronym{simd}{SIMD}{single instruction multiple data}
\newacronym{simt}{SIMT}{single instruction, multiple thread}
\newacronym{fp}{FP}{floating-point}
\newacronym{nan}{NaN}{not a number}

\newacronym{rrm}{RRM}{Radio Resource Management}
\newacronym{ran}{RAN}{Radio Access Network}
\newacronym{rf}{RF}{Radio Frequency}
\newacronym{osi}{OSI}{Open Systems Interconnection}
\newacronym{urllc}{URLLC}{ultra-reliable and low-latency communication}
\newacronym{cots}{COTS}{commercial off-the-shelf}

\newacronym{spr}{SPR}{special-purpose register}
\newacronym{fft}{FFT}{Fast Fourier Transform}
\newacronym{ofdm}{OFDM}{orthogonal frequency division multiplexing}
\newacronym{pla}{PLA}{piecewise linear approximation}
\newacronym{llc}{LLC}{last level cache}

\newacronym{ofm}{OFM}{output feature map}
\newacronym{ifm}{IFM}{input feature map}
\newacronym{fm}{FM}{feature map}

\newacronym{asr}{ASR}{automatic speech recognition}
\newacronym{per}{PER}{Phoneme Error Rate}
\newacronym{lut}{LUT}{Lookup Table}
\newacronym{mcm}{MCM}{multi-chip module}
\newacronym{grs}{GRS}{ground-referenced signaling}
\newacronym{fmc}{FMC}{\Gls{fpga} Mezzanine Card}

\newacronym{ptq}{PTQ}{post-training quantization}
\newacronym{ste}{STE}{straight-through estimator}
\newacronym{qat}{QAT}{quantization-aware training}
\newacronym{inq}{INQ}{incremental network quantization}

\newacronym{rbe}{RBE}{Reconfigurable Binary Engine}
\newacronym{bbq}{BBQ}{Binary Based Quantization}
\newacronym{binconv}{BinConv}{Binary Convolution Engine}
\newacronym{accum}{Accum}{Accumulator Bank}
\newacronym{quant}{Quant}{Quantization Module}

\newacronym{abb}{ABB}{Adaptive Body Biasing}
\newacronym{ocm}{OCM}{On-Chip Monitoring}

\newacronym{fll}{FLL}{frequency-locked loop}
\newacronym{tcdm}{TCDM}{Tightly Coupled Data Memory}
\newacronym{dma}{DMA}{Direct Memory Access}
\newacronym{hwpe}{HWPE}{Hardware Processing Engine}
\newacronym{hci}{HCI}{Heterogeneous Cluster Interconnect}
\newacronym{ip}{IP}{intellectual property}
\newacronym{fsm}{FSM}{finite-state machine}
\newacronym{ai-iot}{AI-IoT}{Artificial Intelligence-enabled Internet-of-Things}

\newacronym{dvs}{DVS}{Dynamic-Voltage Scaling}
\newacronym{dvafs}{DVAFS}{Dynamic-Voltage-Accuracy-Frequency Scaling}
\newacronym{dvas}{DVAS}{Dynamic-Voltage-Accuracy Scaling}
\newacronym{das}{DAS}{Dynamic-Accuracy Scaling}
\newacronym{dadn}{DaDN}{DaDianNao}
\newacronym{ipu}{IPU}{Inner Product Unit}
\newacronym{sipu}{SIPU}{Serial Inner Product Unit}
\newacronym{nm}{NM}{Neuron Memory}
\newacronym{edram}{eDRAM}{embedded DRAM}
\newacronym{mac}{MAC}{multiply-accumulate}
\newacronym{pe}{PE}{Processing Engine}
\newacronym{fusedpe}{Fused-PE}{Fused Processing Engine}
\newacronym{aimc}{AiMC}{Analog in-memory compute} 
\newacronym{imc}{IMC}{in-memory compute} 

\newacronym{lbpe}{LBPE}{Lookup Table-based Processing Engine}
\newacronym{afl}{AFL}{aligned feature map loader}

\newacronym{sb}{SB}{SuperBank}

\newacronym{sdotp}{Sdotp}{sum-of-dot-product}
\newacronym{exsdotp}{ExSdotp}{expanding sum-of-dot-product}
\newacronym{vsum}{Vsum}{vector inner sum}
\newacronym{exvsum}{ExVsum}{expanding vector inner sum}
\newacronym{frep}{FREP}{\gls{fp} repetition instruction}
\newacronym{ssr}{SSR}{stream semantic register}
\newacronym{issr}{ISSR}{indirection stream semantic register}
\newacronym{csr}{CSR}{Control and State Register}

\newacronym{hbi}{HBI}{High-Bandwidth Interconnect}

\newacronym{spi}{SPI}{Serial Peripheral Interface}
\newacronym{uart}{UART}{Universal Asynchronous Receiver Transmitter}
\newacronym{clint}{CLINT}{core local interrupt controller}
\newacronym{plic}{PLIC}{platform-level interrupt controller}

\newacronym{ddr}{DDR}{double data rate}
\newacronym{eda}{EDA}{Electronic Design Automation}

\newacronym[longplural={Scratchpad Memories}]{spm}{SPM}{Scratchpad Memory}
\newacronym{axi}{AXI}{Advanced eXtensible Interface}
\newacronym{cmos}{CMOS}{Complementary Metal-Oxide-Semiconductor}
\newacronym{fpu}{FPU}{Floating Point Unit}
\newacronym{ipc}{IPC}{Instructions Per Cycle}
\newacronym{lsu}{LSU}{Load-Store Unit}
\newacronym{pi}{PI}{programming interface}
\newacronym{pmca}{PMCA}{Programmable Manycore Accelerator}
\newacronym{rtl}{RTL}{Register Transfer Level}
\newacronym{sm}{SM}{Streaming Multiprocessor}
\newacronym{sve}{SVE}{Scalable Vector Extension}
\newacronym{qor}{QoR}{quality of results}
\newacronym[firstplural=core complexes (CCs)]{cc}{CC}{core complex}
\newacronym{hbm}{HBM}{High Bandwidth Memory}
\newacronym{cabf}{CABF}{Cull-and-Aggregate Bottom-Up Floorplanner}
\newacronym{imp}{IMP}{Iterative Merging Packing}
\newacronym{pnr}{PnR}{Place and Route}
\newacronym{tns}{TNS}{Total Negative Slack}
\newacronym{muldiv}{MulDiv}{integer Multiply Divide Unit}
\newacronym{rtwl}{RtWL}{routed wirelength}
\newacronym{cgm}{CGM}{Core Memory Group}
\newacronym{tpc}{TPC}{Texture Processing Cluster}

\newacronym{tlb}{TLB}{translation lookaside buffer}
\newacronym{beol}{BEOL}{backend-of-the-line}
\newacronym{rtc}{RTC}{Real-Time Clock}
\newacronym{pmu}{PMU}{power management unit}
\newacronym{dsp}{DSP}{digital signal processing}
\newacronym{fdsoi}{\mbox{FD-SOI}}{Fully Depleted Silicon-on-Insulator}
\newacronym{nnrf}{\mbox{NN-RF}}{Neural Network Register File}
\newacronym{icache}{I\$}{Instruction Cache}
\newacronym{lic}{LIC}{logarithmic interconnect}
\newacronym{rbeic}{\mbox{RBE-IC}}{RBE interconnect}
\newacronym{lvt}{LVT}{low voltage threshold}
\newacronym{slvt}{SLVT}{super-low voltage threshold}
\newacronym{fbb}{FBB}{forward body biasing}
\newacronym{mcu}{MCU}{micro-controller unit}
\newacronym{pvt}{PVT}{process-voltage-temperature}
\newacronym{mfcc}{MFCC}{mel-frequency cepstrum coefficients}
\newacronym{npu}{NPU}{Neural Processing Unit}
\newacronym{scm}{SCM}{standard-cell memorie}
\newacronym{onnx}{ONNX}{Open Neural Network Exchange}

\begin{document}
%
\title{\textsc{Marsellus}: A Heterogeneous RISC-V AI-IoT End-Node SoC with 2-to-8b DNN Acceleration and 30\%-Boost Adaptive Body Biasing}
%
%
%

\author{Francesco~Conti\orcidlink{0000-0002-7924-933X},~\IEEEmembership{Member,~IEEE,}
        Gianna~Paulin\orcidlink{0000-0002-1310-0911},
        Angelo~Garofalo\orcidlink{0000-0002-7495-6895},~\IEEEmembership{Member,~IEEE,}
        Davide~Rossi\orcidlink{0000-0002-0651-5393},~\IEEEmembership{Member,~IEEE,}
        Alfio~Di~Mauro\orcidlink{0000-0001-6688-1603},~\IEEEmembership{Member,~IEEE,}
        Georg~Rutishauser\orcidlink{0000-0001-8875-7611},~\IEEEmembership{Graduate~Student~Member,~IEEE,}
        Gianmarco~Ottavi\orcidlink{0000-0003-0041-7917},
        Manuel~Eggimann\orcidlink{0000-0001-8395-7585},~\IEEEmembership{Member,~IEEE,}
        Hayate~Okuhara\orcidlink{0000-0003-1582-0100},~\IEEEmembership{Member,~IEEE,}
        and Luca~Benini\orcidlink{0000-0001-8068-3806},~\IEEEmembership{Fellow,~IEEE}%
\thanks{Pre-print manuscript submitted for review to the IEEE Journal of Solid-State Circuits. This work was supported in part by the EU Horizon 2020 project WiPLASH under Grant 863337, in part by the EU Horizon Europe project NeuroSoC under Grant 101070634, in part by the KDT Joint Undertaking project TRISTAN under Grant 101095947, and in part by the Convolve project evaluated by the EU Horizon Europe under Grant 101070374 and supported by the Swiss State Secretariat for Education Research and Innovation under contract number 22.00150. \textit{(Corresponding author: Francesco Conti.)}}%
\thanks{F. Conti, D. Rossi, A. Garofalo, and G. Ottavi are with the Department of Electrical, Electronic, and Information Engineering (DEI), University of Bologna, 40126 Bologna, Italy; e-mail: f.conti@unibo.it.}%
\thanks{G. Paulin, A. Di Mauro, G. Rutishauer, M. Eggimann are with the Integrated Systems Laboratory, ETH Z\"urich, 8092 Z\"urich, Switzerland; e-mail lbenini@iis.ee.ethz.ch.}%
\thanks{H. Okuhara is currently with Department of Electrical and Computer Engineering, National University of Singapore, Singapore, and performed this work while at the University of Bologna.}%
\thanks{L. Benini is with the University of Bologna, 40126 Bologna, Italy, and also with ETH Z\"urich, 8092 Z\"urich, Switzerland.}
}

%
%

\markboth{Post-print accepted for publication by IEEE Journal of Solid-State Circuits}%
{Conti \MakeLowercase{\textit{et al.}}: Marsellus}
%



\maketitle

\begin{abstract}
Emerging Artificial Intelligence-enabled Internet-of-Things (AI-IoT) \glspl{soc} for augmented reality, personalized healthcare, and nano-robotics need to run many diverse tasks within a power envelope of a few tens of mW over a wide range of operating conditions: compute-intensive but strongly quantized \gls{dnn} inference, as well as signal processing and control requiring high-precision floating-point.
We present \textsc{Marsellus}, an all-digital heterogeneous \gls{soc} for AI-IoT end-nodes fabricated in GlobalFoundries 22nm FDX that combines 1) a general-purpose cluster of 16 RISC-V \gls{dsp} cores attuned for the execution of a diverse range of workloads exploiting 4-bit and 2-bit arithmetic extensions  (\texttt{XpulpNN}), combined with fused MAC\&LOAD operations and floating-point support; 2) a 2-8bit \gls{rbe} to accelerate 3$\times$3 and 1$\times$1 (pointwise) convolutions in DNNs; 3) a set of \gls{ocm} blocks connected to an \gls{abb} generator and a hardware control loop, enabling on-the-fly adaptation of transistor threshold voltages.
\marsellus{} achieves up to 180 Gop/s or 3.32 Top/s/W on 2-bit precision arithmetic in software, and up to 637 Gop/s or 12.4 Top/s/W on hardware-accelerated \gls{dnn} layers. 
\end{abstract}


%
\IEEEpeerreviewmaketitle

\glsresetall
\section{Introduction}
\label{sec:intro}

The last few years have witnessed the emergence of a plethora of applications~\cite{li2021survey}, such as augmented reality~\cite{liuAugmentedRealityNext2022,dongSplitNetsDesigningNeural2022,abrashCreatingFutureAugmented2021}, personalized healthcare~\cite{tsinganosDeepLearningEMGbased2023,burrelloBioformersEmbeddingTransformers2022}, and nano-robotics~\cite{vitaleEventdrivenVisionControl2021,oconnellMetaLearningBasedRobustAdaptive2021,niculescuImprovingAutonomousNanoDrones2021}, that require to combine two very distinct sets of characteristics: on the one hand, the low power profile, versatility and flexibility of \gls{iot} endpoint sensor nodes; and on the other hand, the computational power and energy efficiency of hardware accelerators, combined with advanced memory hierarchies, required to enable at-edge inference of \gls{ai} algorithms such as \glspl{dnn}.
AI-IoT \glspl{soc} designed to meet this challenge need to be able to run real-world neural workloads in the range of hundreds of millions of \gls{mac} operations while respecting real-time constraints in the order of milliseconds and staying within an ultra-tight peak power envelope of a few tens of mW to enable long-term operation of battery operated nodes.

To add up to this challenge, AI-IoT applications often combine a primary task employing \glspl{dnn} with other ancillary ones; for instance, on autonomous robots, control-oriented tasks are mixed with \glspl{dnn}~\cite{niculescuImprovingAutonomousNanoDrones2021}, while many audio processing applications combined \glspl{dnn} with \gls{dsp} tasks such as \gls{mfcc}~\cite{fariselliIntegerOnlyApproximatedMFCC2021} computation.
The variety of tasks adds another dimension to the requirements of AI-IoT nodes, which must be capable of quickly ramp-up their performance in a few key computationally intensive kernels, selected at design time; deliver a generally good throughput on other compute-bound tasks; and minimize power consumption in all other states.
Not only the intrinsic workload of such diverse kernels is highly variable, but they also show extremely different characteristics in terms of precision requirements.
In particular, while many signal processing algorithms require high-precision floating-point computations, \glspl{dnn} are generally tolerant to aggressive bit-precision reduction; several techniques for Post-Training Quantization~\cite{nagel2019data} and Quantization-Aware Training~\cite{zhou2017incremental} targeting \glspl{qnn} have been recently proposed.

\begin{figure*}[tb]
    \centering \includegraphics[width=0.95\textwidth]{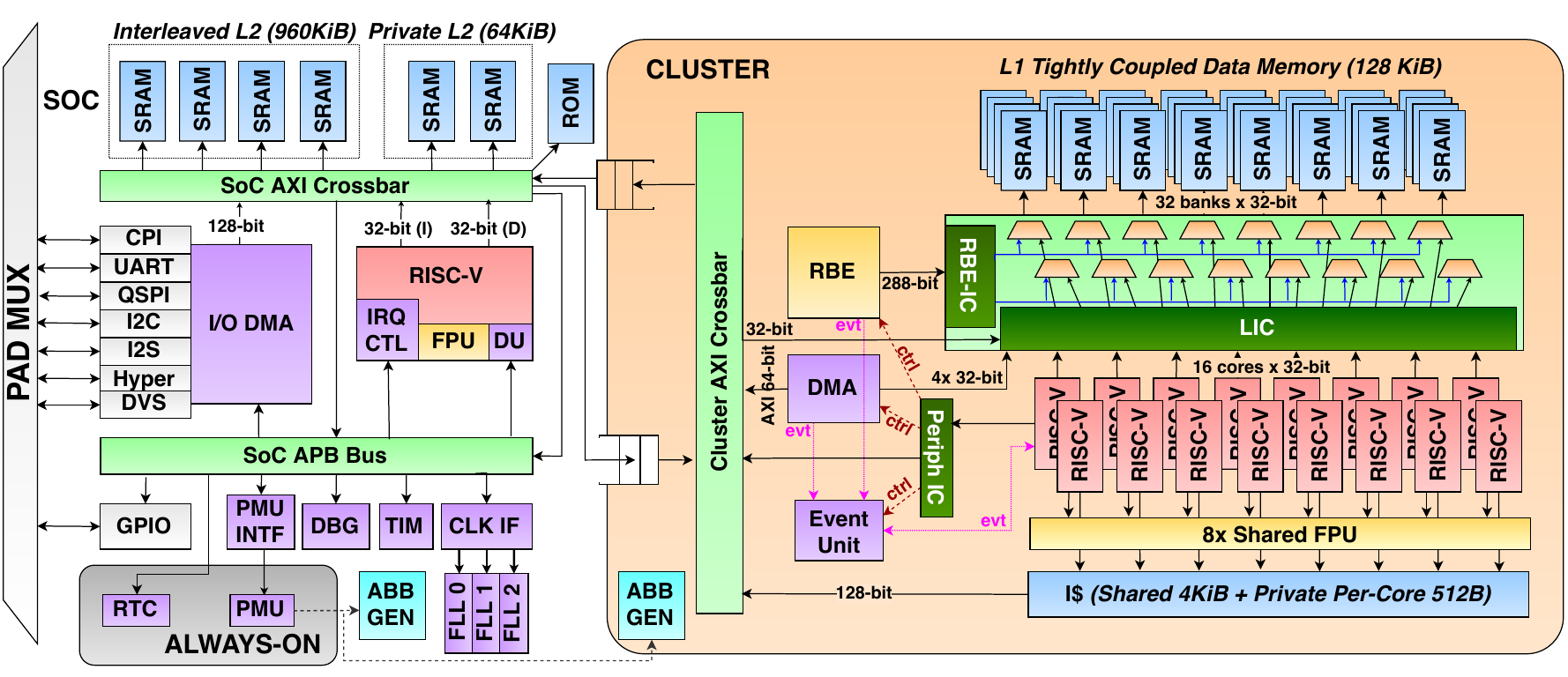}
    \caption{\marsellus{} SoC architecture detailing the three power domains (\textsc{always-on} in light grey, \textsc{soc} in white, \textsc{cluster} in light orange) and their respective main component blocks.}
    \label{fig:fig1_socarchi}
\end{figure*}

State-of-the-Art AI-IoT \gls{soc}'s~{\cite{rossiVegaTenCoreSoC2022,miro-panadesSamurAIVersatileIoT2022,houshmandDIANAEndtoEndHybrid2023,mo12TOPSQuantized2022,jainTinyVersTinyVersatile2023} typically combine a microcontroller, used to marshal data from multiple sensors and connect with other devices, with a hardware acceleration engine to exploit the intrinsic data-level parallelism of compute-intensive \gls{dsp} and \gls{dnn} kernels and the latter's tolerance to bit-precision reduction.
Both fixed-function~\cite{mo12TOPSQuantized2022} and programmable accelerators~\cite{rossiVegaTenCoreSoC2022} based on multi-core clusters have been proposed; however, neither solution is perfect: the former are not flexible to post-fabrication algorithmic changes, e.g., due to the fast pace of development of new \gls{ai} algorithms; the latter cannot deliver enough performance or performance-per-watt to enable breakthrough \gls{ai}-based applications at the edge.
A strategy to sidestep these limitations is to employ aggressive architectural heterogeneity, with heterogeneous acceleration introduced into software-flexible engines at the level of \gls{isa} extensions and coarse-grain hardware accelerators.


\gls{fdsoi} technology~\cite{blagojevicFastFlexiblePositive2016,gomezDesignMethodologyBody2017,moursy35021mm2PVTAware2021} presents a complementary, orthogonal angle of attack to maximize energy efficiency on AI-IoT \gls{soc}'s executing complex applications composed of multiple tasks: the availability of Forward or Reverse Body Biasing techniques, using which it is possible to tune the effective transistor threshold voltage after fabrication to boost performance or improve energy efficiency without scaling the operating frequency, and therefore without performance loss.

In this work, we extend our ISSCC'23 paper~\cite{conti2023marsellus} presenting \marsellus{}, an all-digital AI-IoT end-node heterogeneous \gls{soc} fabricated in GlobalFoundries 22nm FDX technology.
\marsellus{} combines parallel and heterogeneous acceleration with aggressive body-biasing-enabled performance and voltage scalability, leading to state-of-the-art overall performance and efficiency in all application scenarios.
In detail, we introduce the following contributions:
\begin{enumerate}
  \item A general-purpose high-performance cluster of 16\x{} software-programmable RISC-V cores, attuned for execution of a diverse range of DSP workloads exploiting integer (8-bit, 16-bit, 32-bit) and floating-point (16-bit, 32-bit);
  \item \texttt{XpulpNN}, a set of extensions to the RISC-V Instruction Set Architecture (ISA) introducing single-instruction multiple-data execution of low-bitwidth (4-bit, 2-bit) dot-product operations in DNNs, with overlapped MAC\&LOAD operations;
  \item the \rbe{}, a dedicated accelerator for 3$\times$3 and 1$\times$1 (pointwise) convolution layers with a re-configurable datapath to support 2-8bit activation and 2-8bit weight precisions enabling full exploitation of the bitwidth compressibility of \glspl{qnn}.
  \item an \gls{abb} mechanism based on a set of \gls{ocm} blocks and a hardware control loop, enabling automatic on-the-fly adaptation of transistor threshold voltages depending on application requirements.
\end{enumerate}

Our experimental results, measured on a fabricated prototype, show that \marsellus{} delivers leading performance and efficiency on parallelizable software (up to 180 GOPS or up to 3.32TOPS/W with 2-bit precision exploiting the MAC\&LOAD), combined with up to 637 GOPS of performance or up to 12.4 TOPS/W of energy efficiency on key DNN kernels supported by \gls{rbe}: a result comparable to state-of-the-art digital accelerators, with no sacrifice in terms of flexibility and programmability.
Combined with this, the \gls{abb} mechanism provides up to 30\% performance boost in terms of operating frequency even when dropping the operating voltage to save energy.

The rest of this paper is organized as follows.
Section~\ref{sec:architecture} discusses the architecture of the \marsellus{} SoC.
Section~\ref{sec:results} details the experimental setup and results obtained on the SoC.
Section~\ref{sec:dnn} discusses mapping of DNNs on \marsellus{}.
Section~\ref{sec:soa} performs a detailed comparison of our work with some of the main related works in the State-of-the-Art.
Section~\ref{sec:conclusion} draws conclusions.

\section{Marsellus SoC Architecture}
\label{sec:architecture}

Fig.~\ref{fig:fig1_socarchi} details the architecture of the \marsellus{} \gls{soc}.
\marsellus{} is designed for the diverse needs of IoT devices, typically featuring a microcontroller unit and AI accelerators, focusing on parallel and heterogeneous computing.
Accordingly, \marsellus{} is organized in three power domains (\textsc{always-on} in light grey, \textsc{soc} in white, \textsc{cluster} in light orange), the latter two of which correspond to an advanced microcontroller and to a heterogeneous accelerator, respectively.
\textsc{soc} and \textsc{cluster} constitute also distinct clock domains, communicating through a set of dual-clock AXI FIFOs.

The \textsc{soc} implements an advanced microcontroller based on a RISC-V \texttt{RV32IMCFXpulp} core~\cite{Gautschi2017}, based on an in-order 4-stage pipeline designed to achieve high \gls{ipc} in general-purpose and arithmetics-oriented applications.
The \textsc{soc} core features a \gls{fpu} supporting the full RISC-V floating-point \gls{isa} extension and is augmented with the DSP-oriented \texttt{Xpulp} extension, which implements two nested hardware loops, post-increment load/store instructions,  fused integer \gls{mac} instructions, and dot-product instructions for 16-bit and 8-bit data.
The \textsc{soc} includes also a large L2 \gls{sram}-based scratchpad memory divided in a 4-bank word-interleaved section (\SI{960}{\kibi\byte}) and a private bank-interleaved section (\SI{64}{\kibi\byte}).
Both instructions and data (stacks, heap) are managed by the core and can be allocated to either section; the boot code is allocated on a small (\SI{8}{\kibi\byte}) boot ROM.
All memories are accessed from the \textsc{soc} core by means of a 64-bit AXI4 crossbar.

The \textsc{soc} includes an I/O \gls{dma} controller~\cite{pullini2017mudma} capable of marshaling data to/from the L2 memory with up to 128~bit/cycle bandwidth, from/to several I/O interfaces (including QSPI, I2C, I2S, Cypress' HyperRAM protocol for external memory, and an interface for DVS cameras~\cite{li13210410mmPixel2019}).
The \textsc{soc} is completed by a set of peripherals accessed via an APB peripheral bus: GPIOs, debug unit, and timers.
Finally, the aforementioned \textsc{always-on} island contains a \gls{rtc} and a \gls{pmu}, which controls the \gls{abb} generator and enables coarse-grain power gating of the \textsc{soc} and \textsc{cluster} domains. It also contains three \glspl{fll} to generate separate clocks for the \textsc{soc} core \& memories, for \textsc{soc} peripherals and for the \textsc{cluster} domain.

The \textsc{cluster} is a separate power and frequency island hosting 16 identical RISC-V \gls{dsp} cores implementing the \texttt{RV32IMFCXpulpnn} \gls{isa}.
Similarly to the \textsc{soc} core, the \textsc{cluster} ones are based on the RI5CY baseline architecture; however, they are further augmented with the \texttt{Xpulpnn} \gls{isa} extension.
\texttt{Xpulpnn} is a superset of \texttt{Xpulp} that introduces support for sub-byte (2-bit, 4-bit, 8-bit) symmetric precision dot-product instructions and a fused MAC\&LOAD mechanism, relying on a dedicated \gls{nnrf}, that enables the cores to achieve near-100\% MAC unit utilization during the execution of linear algebra kernels.
The architecture of the \textsc{Marsellus} cluster RISC-V cores is discussed in detail in Sec.~\ref{sec:macandload}.
The 16 cores share a hierarchical \gls{icache}  composed of a \SI{4}{\kibi\byte} of 4-way associative, 128-bit/line shared cache (L1.5) common between all cores, with the addition of smaller \SI{512}{\byte} L1 private per-core caches~\cite{jie2020energy}.
The L1.5 shared \gls{icache} employes multi-port (MP) memories to remove direct critical paths between different L1 \glspl{icache} and enable scaling the cluster to 16 cores. The L1 private per-core \glspl{icache} minimize the path between the RISC-V core prefetchers and the L1.5 \gls{icache}, enabling higher clock speed. Both \gls{icache} levels are realized with \glspl{scm} to enable the MP architecture and improve the overall energy efficiency compared with regular \gls{sram} cuts.
The cores also share and 8 \glspl{fpu} with support for IEEE 32-bit float, IEEE 16-bit float, and BF16 formats, for efficient support of floating-point \gls{dsp} applications~\cite{montagnaLowPowerTransprecisionFloatingPoint2022}.

Together with the RISC-V cores, the main computational element of the \textsc{cluster} is the \gls{rbe}: an accelerator for \gls{dnn} convolution layers with a unified datapath that can be runtime-configured in different modes (3$\times$3 and 1$\times$1 convolutions) and activation/weight precisions (asymmetric 2--8 bits).
Other layers (e.g., fully-connected, 3\x{}3 depth-wise convolution) can be implemented as corner cases of the two natively supported modes, whereas unsupported layers are executed on the \textsc{cluster} RISC-V cores.
A 64-bit/cycle read, 64-bit/cycle write \gls{dma} engine can be used to marshal data between the L2 memory in the \textsc{soc} and the \textsc{cluster}, through a \textsc{cluster}-level AXI crossbar connected to the \textsc{soc} through dual-clock AXI FIFOs.
Cores, \gls{dma} engine and \gls{rbe} share at L1 the same \SI{128}{\kibi\byte} of \gls{sram} \gls{tcdm}.
The \gls{tcdm} is organized in 32 word-interleaved banks to provide high-bandwidth parallel access to all the traffic generators; access is delivered via a 0-wait-state, 928-bit/cycle aggregate bandwidth \textsc{cluster} interconnect.
The interconnect is organized hierarchically and split in two branches; the \gls{lic} branch is a fully combinational crossbar to route \& arbitrate accesses from the 16 cores, the \gls{dma} and a further 32-bit \textsc{soc} port towards the \gls{tcdm} banks; the \gls{rbeic} branch routes RBE accesses, which are always contiguous, towards \gls{tcdm} banks with no  bank-wise arbitration.
A set of bank-level multiplexers are employed to grant access to the \gls{lic} or \gls{rbeic} branch, utilizing a round-robin rotation scheme to avoid starvation.

The \textsc{cluster} includes also an Event Unit to enable high-performance parallel programming synchronization primitives (barriers, critical sections, etc.) between the 16 cores, as well as for fast communication of end-of-job events from the \gls{dma} and the \gls{rbe}.
Finally, the cluster interconnect also includes a secondary 32-bit peripheral interconnect that is used for the configuration of \gls{rbe}, the \gls{dma}, and the Event Unit.
All components of the \textsc{Marsellus} \textsc{cluster} are tightly coupled and can not be power-gated at a fine grain as, e.g., the modules in Jain~\textit{et~al.}~\cite{jainTinyVersTinyVersatile2023}.
The specific bus widths and memory sizes utilized in the \textsc{cluster} were chosen to enable the target applications of \marsellus{} to be run in a compute-bound scenario in most cases, supporting both RISC-V and \gls{rbe}-accelerated computation.

\begin{figure*}[tb]
    \centering \includegraphics[width=0.95\textwidth]{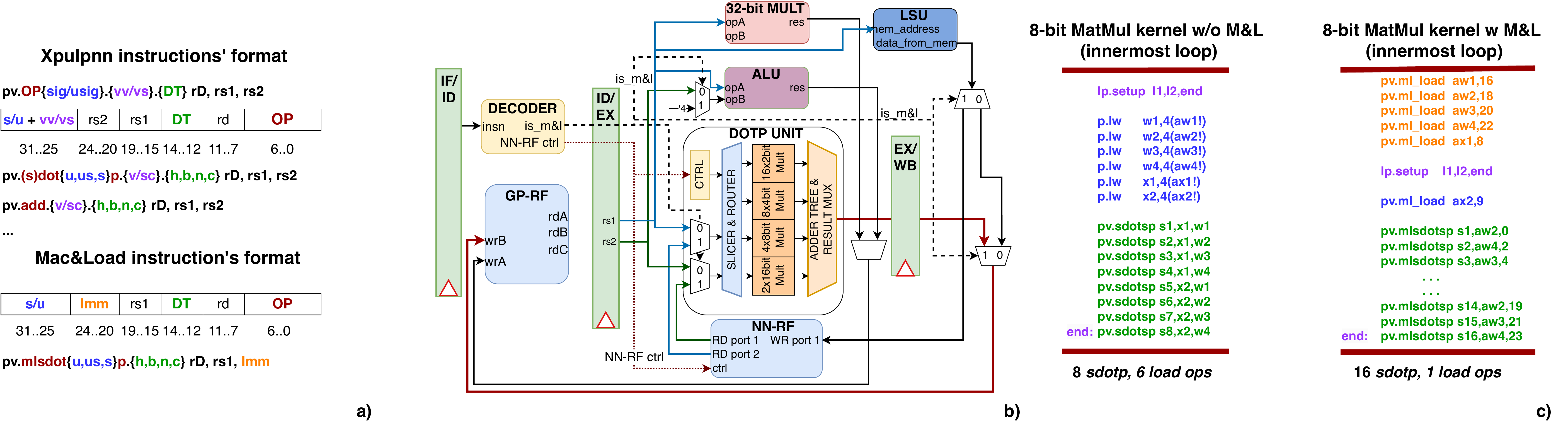}
    \caption{a) Encoding formats of the \texttt{Xpulpnn} instructions, including the MAC\&LOAD. b) Modified RI5CY pipeline to implement the \texttt{Xpulpnn} extension with \textit{i}) low-bitwidth dot-product units and \textit{ii}) additional \gls{nnrf} for MAC\&LOAD support. c) Assembly code snippets of innermost loops of 8-bit symmetric matrix multiplication kernels (MatMul), implemented without (left-sided code) and with MAC\&LOAD. }
    \label{fig:fig3_xpulpnn}
\end{figure*}


\subsection{\texttt{Xpulpnn} extensions and MAC\&LOAD}
\label{sec:macandload}

\subsubsection{Operating principle}
As previously introduced, the main innovation of Marsellus that we present at the core level is the design, at architectural and micro-architectural level, of a set of ISA instructions, namely \texttt{XpulpNN}, aiming to boost the performance and efficiency of reduced-precision integer DSP and linear algebra kernels. 

We build such extensions on top of \texttt{Xpulp} extending the support in the ISA of the RISC-V processor to packed-SIMD operations performed on vectors of nibble and crumb data types, i.e. of 4-bits and 2-bits respectively. The core of \texttt{Xpulpnn} consists of dot-product (\textit{dotp}) operations, including the MAC-equivalent sum-of-dot-product (\textit{sdotp}), executed in various formats: the two inputs can be both vectors (\textit{vv}) or one vector and the other a scalar value replicate in each vector element (\textit{vs}); the vectors can be interpreted as both signed (\textit{s}), both unsigned (\textit{u}) or the first unsigned and the second signed (\textit{us}) (and viceversa); the third scalar operand and the accumulator feature always 32-bit precision and can be either signed or unsigned. \texttt{Xpulpnn} includes also nibble and crumb packed-SIMD ALU operations, such as vector addition, subtraction, maximum, minimum, shuffling, and other instructions for bit-manipulation at the granularity of the vector elements. 

To significantly improve the utilization of the processor’s pipeline on regular kernels like the matrix-multiplication, we design a fused MAC\&LOAD instruction, which applies to all the \textit{dotp} SIMD formats supported by \texttt{Xpulpnn}, that collapses one packed-SIMD \textit{dopt} and one post-modified load operations into the same instruction. Since the data-path activated by the \textit{dotp} does not interfere with the Load-Store Unit (LSU), the two units can run in parallel without requiring complex logic to control the instruction flow. 

\begin{figure*}[tb]
    \centering \includegraphics[width=0.95\textwidth]{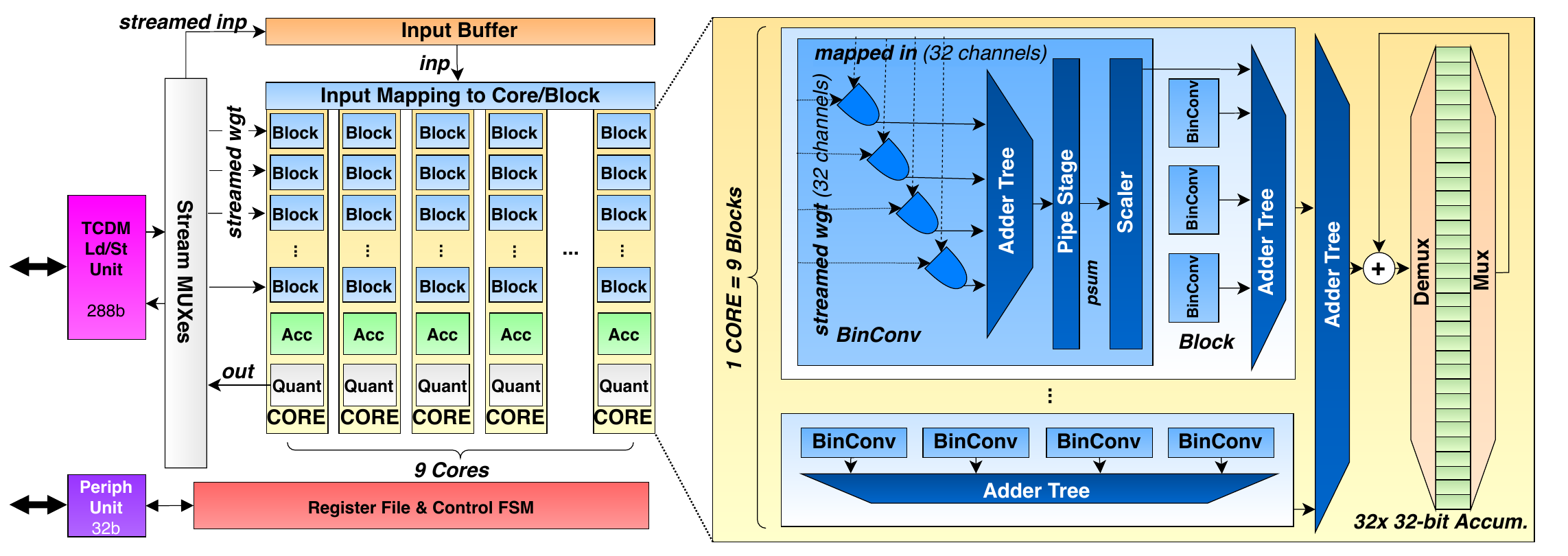}
    \caption{\gls{rbe} microarchitecture (left), showing the 9 Cores divided in 9 Blocks each; and detail of a \gls{rbe} Core microarchitecture (right).}
    \label{fig:fig2_rbe_archi}
\end{figure*}

\subsubsection{Microarchitecture}
As shown in Fig.~\ref{fig:fig3_xpulpnn} b), to implement the \texttt{Xpulpnn} instructions, we modify the micro-architecture of the baseline RI5CY core as follows: we extend the DOTP unit of RI5CY, which consists of two sets of multipliers supporting 16- and 8-bit packed-SIMD \textit{dotp} operations, with two additional multipliers islands for nibble and crumb operands. We choose to replicate the multipliers islands for different precisions to minimize the impact of \texttt{Xpulpnn} on the critical path of the core: extra logic to manipulate and reshape the operands before feeding the single multi-precision island would increase the critical path of the core. This helps improving efficiency because the multi-branch datapath uses operand isolation to remove spurious switching activity.  At the same time, we enhance the ALU with the support for operations on nibble and crumb operands and we integrate the new instructions into the decoder of the pipeline.

The design of the MAC\&LOAD is more complex and it is the result of an architecture/micro-architecture codesign to maximize the efficiency of the instruction and minimize the hardware costs to implement it. In reference to the Fig.~\ref{fig:fig3_xpulpnn} a) and b) that shows the encoding of the instruction and the micro-architecture of the core, the MAC\&LOAD works as follows: it fetches the operands of the \textit{dotp} operation from a dedicated register file, namely NN-RF, containing 6 32-bit SIMD vector registers, addressed by a 5-bit immediate field of the instruction; the accumulator resides in the GP-RF of the core instead and it is updated once the \textit{dopt} operation is completed, in the EX stage. One of the addressed NN-RF register can be refreshed with a new data from the memory. In such case, one of the two most significant bits of the immediate is set (note that, by construction, only one of the registers can be refreshed with one instruction, since the core's pipeline features a single 32-bit LSU data port towards the memory) and the LSU of the core receives from the GP-RF the pointer for the memory access. Such pointer is then incremented by one word in the ALU and stored back into the GP-RF in the EX stage, while the data fetched from the memory, available in the WB stage of the pipeline, is directly routed to the write port of the NN-RF.

The operating mechanism of the MAC\&LOAD instruction, with the dedicated NN-RF, has three important advantages: first, it avoids to add a costly write port to the GP-RF (which features only two write ports in the baseline RI5CY core), otherwise necessary because the MAC\&LOAD needs to store three results: the dopt result, the updated pointer and the new data fetched from the memory; second, we can directly control how long an operand will reside in the NN-RF without being constrained by the compiler scheduler, allowing more flexibility for data-reuse at register file level, particularly effective to reduce memory traffic and increase the energy efficiency of compute- and memory-intensive kernels (especially critical in the context of multi-core systems, where memory conflicts reduce performance and energy efficiency); third, fetching the \textit{dotp} operands from the NN-RF leaves more room in the GP-RF to store more accumulators: exploited in combination with data reuse techniques, it increases the number of outputs produced in the innermost loops.

Each \texttt{Xpulpnn} RISC-V core in the cluster of Marsellus costs 78kGE, with a total overhead of 17.5\% compared to the baseline RI5CY core, due to the additional multiplier islands in the DOTP unit of RI5CY, the extended ALU and the decoding stage. The extra hardware cost of the MAC\&LOAD is for the NN-RF,  the logic to distribute the operands to the DOTP unit from the NN-RF and to route the data from the LSU to the NN-RF write port. From a power consumption perspective, to avoid unnecessary switching activity when the MAC\&LOAD and/or the classic \textit{dotp} SIMD operations are not executed, we perform operand isolation on critical wires and apply clock-gating to the NN-RF and the DOTP unit, limiting the power overhead of the core on general-purpose applications to $\sim3\%$. 

\subsubsection{Compiler support}
To ease the exploitation of the proposed ISA extension, we integrate the machine-level description of the \texttt{Xpulpnn} operations into the GCC compiler, allowing the programmer to infer these instructions in the C application code through explicit invocation of built-in functions. Compared to inline assembly, this approach enables optimization passes by the compiler back-end that maximizes reuse of operands and efficiently schedules the instruction flow. We provide also a set of optimized C routines, based on the \texttt{Xpulpnn} ISA, for QNN and linear algebra kernels, publicly available under open-source permissive licence\footnote{https://github.com/pulp-platform/pulp-nn-mixed/tree/main/XpulpNN}.

\subsection{Reconfigurable Binary Engine (RBE)}
\label{sec:rbe}


\subsubsection{Operating principle}

As previously introduced, the \gls{rbe} is a hardware accelerator targeting \gls{dnn} convolution layers with a unified datapath that can be reconfigured at runtime in two different modes of operation (3$\times$3 and 1$\times$1 convolutional layers) and activation/weight precisions (asymmetric between 2 and 8 bits, including non-power-of-two bitwidths).
Considering $W$-bit weights, $I$-bit inputs, and $O$-bit outputs, \gls{rbe} splits each $W\times I$-bit product into $W\times I$ distinct single-bit contributions, which are then allocated partially in parallel on different 1-bit \gls{mac} units, and partially serialized in time.

Considering for simplicity a 1$\times$1 convolution, in RBE weights and inputs are decomposed in binary contributions, $\mathbf{wgt}$ and $\mathbf{inp}$, respectively, and the convolution operation is performed in the binary domain accumulating scaled partial sum in a 32-bit accumulator:
\begin{equation}
    \mathbf{acc}_{[h, w, k_{out}]} = \sum_{i=0}^W\sum_{j=0}^I\sum_{k_{in}} 2^{i+j} \cdot \mathbf{wgt}_{[k_{out},k_{in},i]} \wedge \mathbf{inp}_{[h, w, k_{in}, j]}
    \label{eq1}
\end{equation}
where $\wedge$ is a logical AND operation and the multiplication by $2^{i+j}$ is a left-shift.
The 3$\times$3 case is analogous, with a further summation over 9 filter contributions.

After complete accumulation, the value of $\mathbf{acc}$ is normalized and right-shifted:
\begin{equation}
    \mathbf{out}_{[h, w, k_{out}]} = \left( \mathbf{scale}_{[k_{out}]}\cdot\mathbf{acc}_{[h, w, k_{out}]} + \mathbf{bias}_{[k_{out}]} \right) \gg \mathbf{S}
    \label{eq2}
\end{equation}

Implementing convolutions by means of AND, left-shift, and add operations enables finely-controlled mixed precision computation on the same underlying hardware.
For example, a convolution with $W=3$, $I=4$, and $O=2$ can be implemented on the same hardware of one using only 8-bit data representation by tuning shift factors and loop iterations.
The \gls{rbe} microarchitecture implements Eqs.~\ref{eq1} and \ref{eq2} by tiling the inner accumulation and outer loops, and executing part of the inner tile loops on a hierarchical architecture that is discussed in the following.



\begin{figure*}[tb]
    \centering \includegraphics[width=0.95\textwidth]{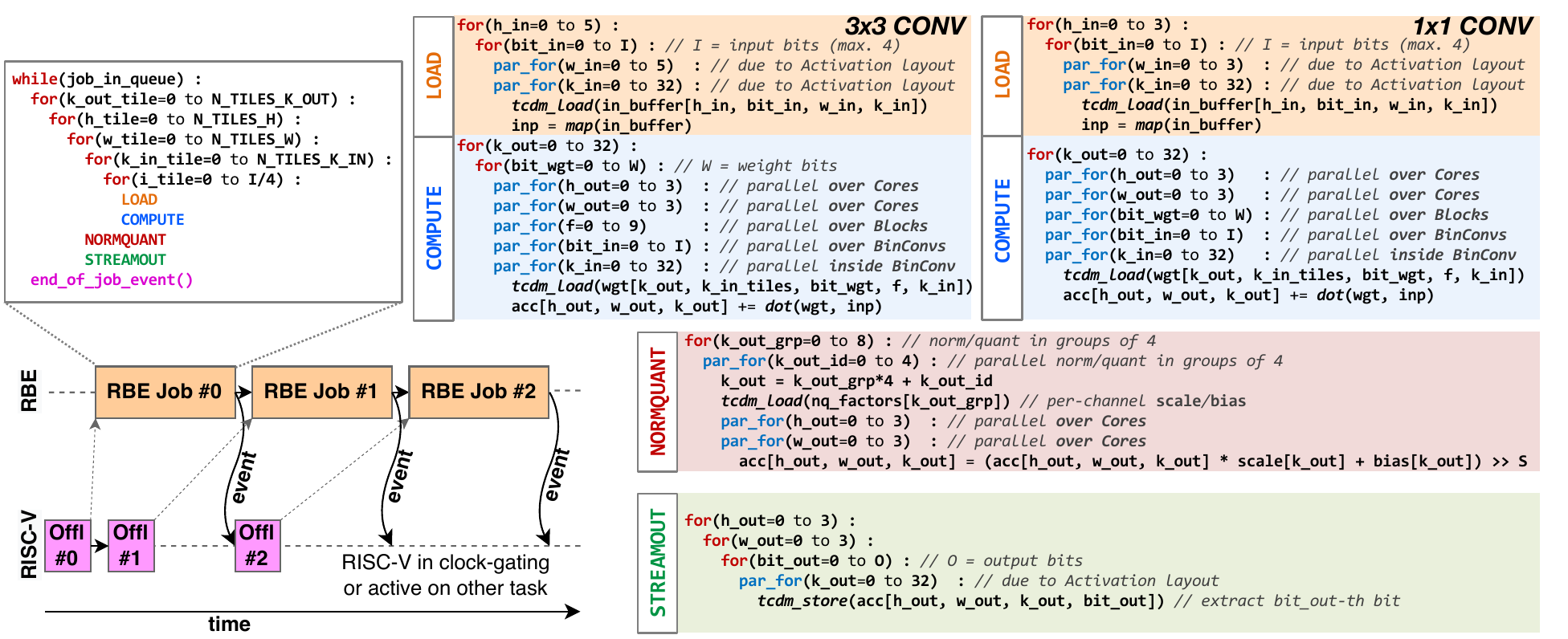}
    \caption{\gls{rbe} execution flow, with multiple jobs offloaded and detail of the dataflow loop nests implemented in the accelerator for 3$\times$3 and 1$\times$1 convolutions (\texttt{\bfseries \color{RedOrange} LOAD} and \texttt{\bfseries \color{RoyalBlue} COMPUTE} phases), normalization/quantization (\texttt{\bfseries \color{BrickRed} NORMQUANT}) and stream-out of outputs (\texttt{\bfseries \color{ForestGreen} STREAMOUT}).}
    \label{fig:fig7_rbe_loopnest}
\end{figure*}

\subsubsection{Microarchitecture}
Fig.~\ref{fig:fig2_rbe_archi} introduces the architecture of \gls{rbe}, which is based on the open-source \gls{hwpe}\footnote{\url{https://hwpe-doc.readthedocs.io}} architectural template and set of \gls{ip} elements.
Following this template, \gls{rbe} is divided into three parts: \textit{controller}, \textit{streamer}, and \textit{datapath} (or engine).
The \gls{rbe} controller consists of a Peripheral Unit, connected as a target to the \textsc{cluster} peripheral interconnect, a latch-based dual-context register file, and a hierarchical \gls{fsm} that controls the overall behavior of a job offloaded to the \gls{rbe}.
To simplify the datapath control, in particular, the tiled loop nests that implement convolutional layers, part of the \gls{fsm} is realized using a \gls{sw} configurable \emph{uloop}, i.e., a tiny microcoded loop processor~\cite{Conti2018a}.

The streamer is composed of a 288-bit wide \gls{tcdm} Load/Store Unit, realized using standard open-source \gls{hwpe} \gls{ip}s; this unit accesses data in the \gls{tcdm} memory and converts it into an internal streaming representation using a simple ready/valid handshake protocol to enable latency insensitiveness of the inner datapath.
The streamer uses a three-dimensional strided address generator and is capable of linearizing any 3D strided memory access pattern into a stream.
A set of multiplexers and de-multiplexers helps the streamer to address the incoming memory stream towards the correct consumer in the \textit{datapath} (or vice versa for outgoing streams).


The \gls{rbe} datapath exploits an output-stationary and partially bit-serial dataflow with reuse over the spatial dimensions mapped onto a set of 9 Cores, each one working on the receptive field of one pixel on 32 channels in the output space.
\gls{rbe} is organized hierarchically into a grid of 9\x{}9 = 81 Blocks, arranged in 9 Cores of 9 Blocks each.
Each Block includes 4\x{} \glspl{binconv}, and each \gls{binconv} performs a 32\x{}32 1-bit dot-product per cycle (using AND gates, and achieving 32 binary \gls{mac}/cycle).
The result of the dot-product reduction is then scaled by the required power-of-two value (according to the operating mode) with the help of small dynamic shifters.
The scaled results of the four \glspl{binconv} are accumulated at Block level, and the cumulative results of all Blocks within a single Core are
accumulated and stored in one of 32\x{} 32-bit latch-based \glspl{accum} in each core.
After the accumulation is complete (i.e., all reduction dimensions in the convolutional layer have been fully computed), a Quantizer module in each Core is used to perform ReLU activation and reduce the 32-bit accumulators into $O$ bits.
In this way, \gls{rbe} supports mixed-precision \glspl{dnn} with $I$, $O$, $W$ arbitrarily set to any value in 2-8 bits.




\subsubsection{Data layout}
To implement the principle of Eqs.~\ref{eq1} and \ref{eq2} on top of the \gls{rbe} microarchitecture, we introduce a specialized data layout in \gls{tcdm} for both weights and activations, which exposes the same parallelism that is used inside the accelerator, by swapping the ``bit index'' dimension with a (tiled) channel dimension.
Specifically, input activation bitstreams are stored in memory in the \textit{(H, W, K/32, I, 32)} format, to align with the BinConv parallelism of 32 channels.
Similarly, output bitstreams are stored in the \textit{(H, W, K/32, O, 32)} format.
Weights are stored in such a way that they can be directly streamed from memory into \gls{rbe} without marshaling.
For 3$\times$3 convolutions, we employ a \textit{(Kout, Kin/32, W, 9, 32)} format, where the two innermost dimensions are aligned with the number of Blocks per Core and with the parallelism of the BinConvs, respectively.
For 1$\times$1 convolutions, we use \textit{(Kout, Kin/32, W, 32)}.

\subsubsection{Execution flow}
RISC-V cores can enqueue up to 2 jobs in the \gls{rbe} register file; whenever the accelerator is free, it will start executing the oldest offloaded job and emit an event to synchronize with the core at the end of each job execution.
Each job performs a complete convolutional layer implementing Eqs.~\ref{eq1} and \ref{eq2}.
Dimensions not aligned with the size of the \gls{rbe} accelerator are tiled and executed sequentially by the control FSM, using the embedded \textit{uloop} unit to implement deeply nested loops with minimal overhead.
Fig.~\ref{fig:fig7_rbe_loopnest} specifies in detail the complete execution flow of \gls{rbe} in the form of nested temporal loops (\texttt{\bfseries \color{BrickRed}for}) and parallel execution (\texttt{\bfseries \color{NavyBlue}par\_for}); the core of the execution is composed by the \texttt{\bfseries \color{RedOrange} LOAD} and \texttt{\bfseries \color{RoyalBlue} COMPUTE} states, which operate differently in the two 1$\times$1 and 3$\times$3 operating modes.
After accumulation is complete, the \gls{rbe} performs normalization (\texttt{\bfseries \color{BrickRed} NORMQUANT}) and stream-out of outputs (\texttt{\bfseries \color{ForestGreen} STREAMOUT}).


For the 3\x{}3 convolutional mode, \rbe{} loads an input patch of 4-bits (less if $I<4$) of 32 channels of 5\x{}5 pixels into the input buffer.
Each of the 9 Cores works on the receptive field of a single pixel on 32 channels in the output space.
The 32 channels of the 3\x{}3 filters are unrolled over the 9 Blocks of each Core, broadcasted to all 9 Cores, and bit-serialized in time.
With the output-stationary flow, the partial output results are stored in the latch-based \glspl{accum} while new input patches are loaded if $I>4$, or the \kin{}$>32$.
For the 1\x{}1 convolutional mode, the streamers load a smaller patch of up to 4-bits (less if $I<4$) of 32 channels of 5\x{}5 pixels into the input buffer.
The individual \textit{W} bits of the weight are now mapped in a bit-parallel fashion on the Blocks of a Core while still being broadcasted to all 9 Cores.
The last Block of each Core remains, therefore, unused and clock-gated in this operation mode.
Overall, running jobs with lower $W$ results in a faster computation time for the 3\x{}3
operation mode and a reduced acceleration utilization for 1\x{}1 operation mode.
The \gls{rbe}, with its total of 10368 AND gates used as single-bit multipliers, achieves peak throughput (1610 operations/cycle in the \texttt{\bfseries\color{RoyalBlue}COMPUTE} state) in the 3\x{}3 mode with \textit{W}=2, \textit{I}=2 or 4, \textit{O}=2 or 4.
We refer to Section~\ref{sec:rbe_perf} for a complete analysis of the \gls{rbe} performance in several operating modes.

\subsection{On-Chip Monitors and Adaptive Body Biasing}\label{sec:abb}

\begin{figure}[tb]%
    \centering
    \includegraphics[width=\columnwidth]{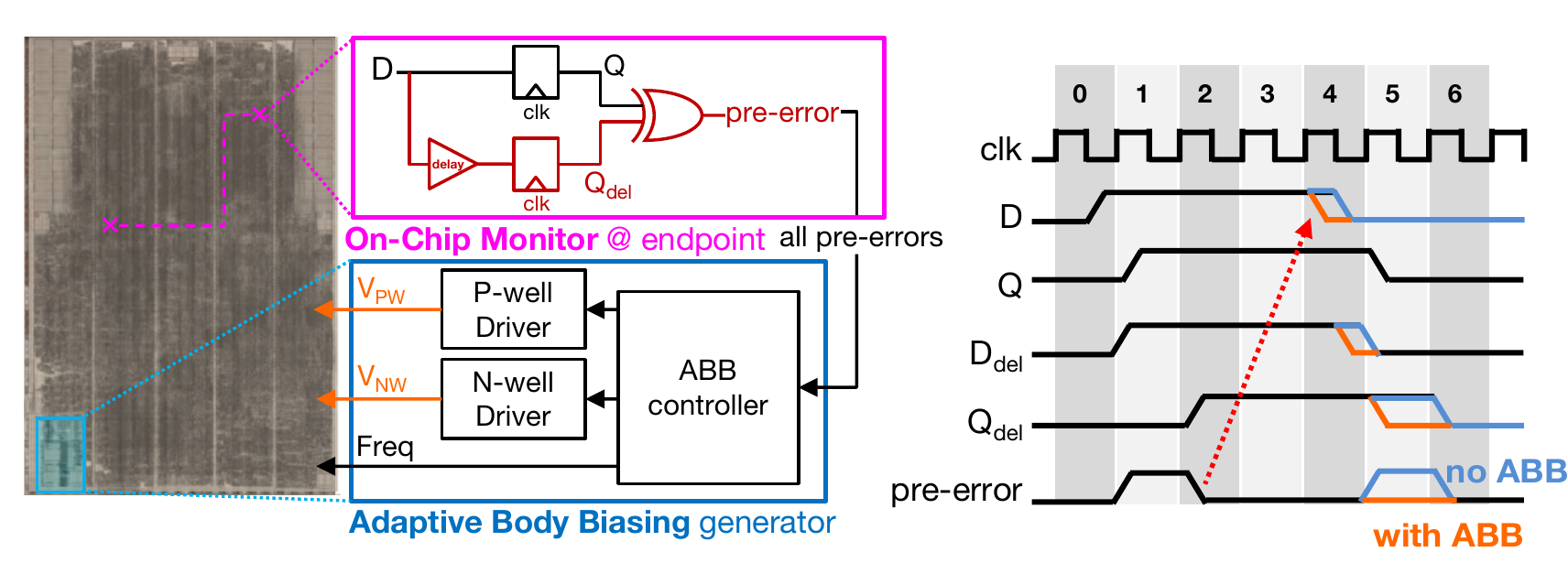}
    \caption{(left) Adaptive Body Biasing (ABB) mechanism implemented in \marsellus{} with On-Chip Monitors detecting pre-errors at 1\% of timing critical endpoints and ABB generator; (right) an example of pre-error detection and ABB generation.}
    \label{fig:fig10_abb_explanation}
\end{figure}

\begin{figure}[tb]
    \centering \includegraphics[width=0.95\columnwidth]{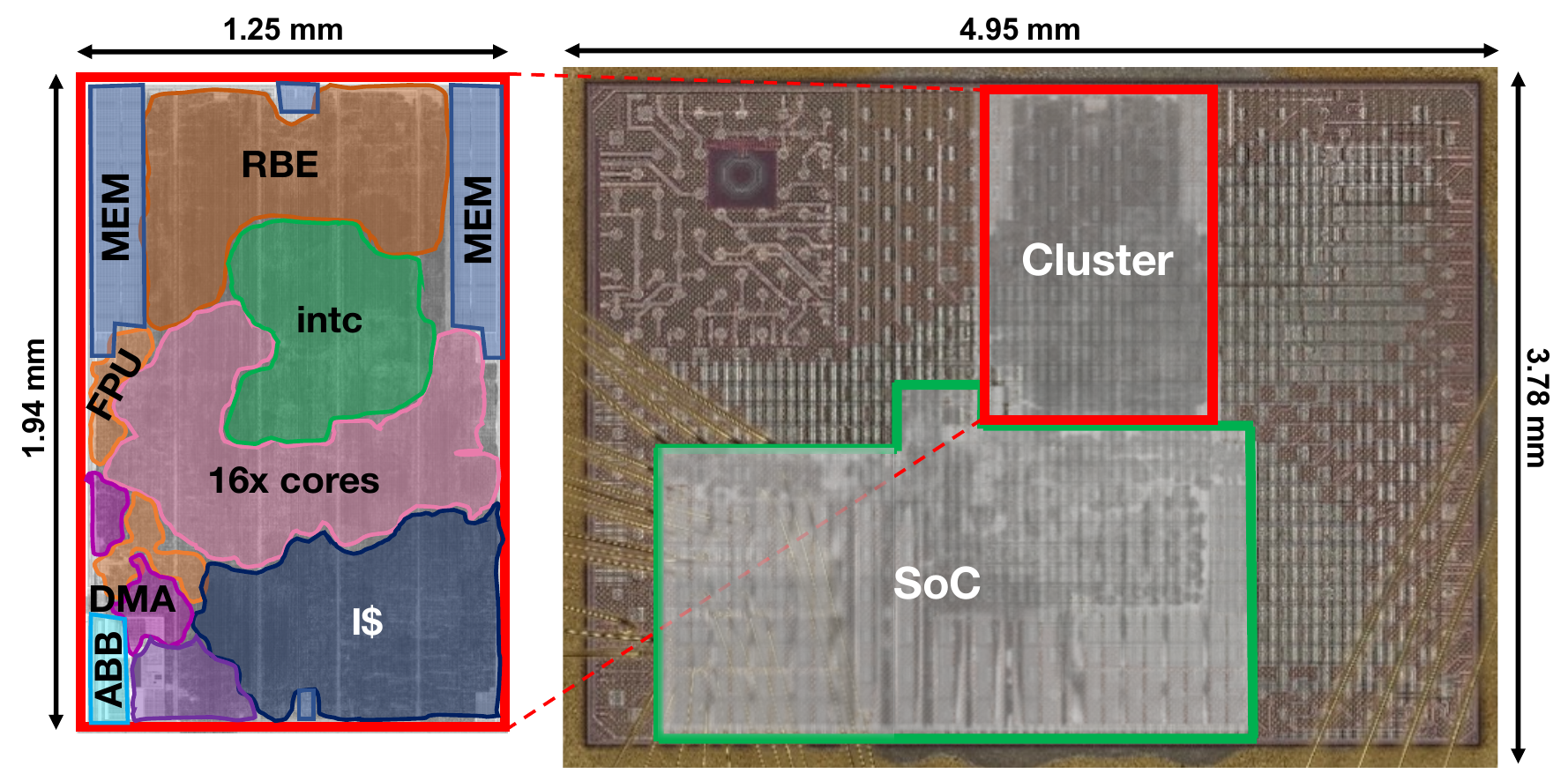}
    \caption{Microphotograph of the \marsellus{} prototype fabricated in GlobalFoundries 22FDX technology.}
    \label{fig:fig0_micrograph_recap}
\end{figure}

To further improve energy efficiency in computationally demanding applications beyond the described architectural innovation, \marsellus{} introduces a dynamic Adaptive Body Biasing (\gls{abb}) mechanism, based on the circuit introduced in Moursy~et~al.~\cite{moursy35021mm2PVTAware2021}, whose operation is shown in Fig.~\ref{fig:fig10_abb_explanation}.
The cluster ABB generator incurs in a very small area (\SI{0.039}{\square\milli\meter}) and power (+0.4\%) overhead in exchange for significant performance and energy efficiency gains, that are discussed in the following.
The \gls{abb} mechanism is based on the empirical observation that after timing closure, a minority of combinational register-to-register paths in a system will remain near-critical, i.e., have a very small positive slack, due to their length (number of gates, wire length).
Reducing the supply voltage without scaling frequency (or, conversely, increasing frequency but not voltage) will result in timing failures that, with a high probability, will hit one or more of these near-critical paths.

At \marsellus{}' signoff, the 1\% of the register-to-register path endpoints with smallest positive slack (i.e., nearest to being critical) were selected and augmented with On-Chip Monitors.
\glspl{ocm} work by pairing the endpoint register with a shadow copy, which is fed with a delayed version of the endpoint register input.
XOR-ing the outputs of the functional and shadow registers, \glspl{ocm} detect whether an endpoint is close to becoming timing-critical (e.g., when the \gls{soc} is under-volted or over-clocked), and raise a \textit{pre-error} signal that is propagated to the \gls{abb} generator (Fig.~\ref{fig:fig10_abb_explanation}).

The \gls{abb} generator collects all pre-error signals and, depending on its configuration, its internal hardware loop can react by directly tuning the \gls{soc}'s N-well and P-well biasing voltages to increase \gls{fbb}, which in turn reduces the logic's voltage threshold and hence, all propagation delays.
Conversely, if it does not detect any pre-error the generator will progressively reduce the body biasing voltage, thereby raising the devices' thresholds, to save power.
By properly calibrating the pre-error delay margins detected by \glspl{ocm}, the \gls{abb} effectively dynamically trims the setup margins of flip-flops, enabling a higher frequency operation at the same voltage.

Fig.~\ref{fig:fig10_abb_explanation} (right) shows an example.
In the absence of \gls{abb} (blue line), a pre-error condition is ignored, and after a few cycles, a real error arises due to a setup time violation.
Conversely, with \gls{abb}, the pre-error condition is detected, and the generator reacts by increasing forward body biasing,  lowering thresholds and preventing the error condition thanks to the consequent speed-up.

\section{SoC Measurements}
\label{sec:results}

\subsection{Area, frequency and power}

\begin{figure}[tb]
  \centering
  \resizebox{\columnwidth}{!}{%
\def\innerradius{0cm}
\def\innerradiuss{0.5cm}
\def\outerradius{1.6cm}
\def\rotate{0}

\newcommand{\wheelchart}[1]{
    \pgfmathsetmacro{\totalnum}{0}
    \foreach \value/\colour/\name in {#1} {
        \pgfmathparse{\value+\totalnum}
        \global\let\totalnum=\pgfmathresult
    }

    \begin{tikzpicture}

      \pgfmathsetmacro{\wheelwidth}{\outerradius-\innerradiuss}
      \pgfmathsetmacro{\midradius}{(\outerradius+\innerradiuss)/2}

      \begin{scope}[rotate=\rotate]

      \pgfmathsetmacro{\cumnum}{0}
      \foreach \value/\colour/\name in {#1} {
            \pgfmathsetmacro{\newcumnum}{\cumnum + \value/\totalnum*360}

            \pgfmathsetmacro{\percentage}{\value/\totalnum*100}
            \pgfmathsetmacro{\midangle}{-(\cumnum+\newcumnum)/2}
            \pgfmathsetmacro{\midanglerot}{\midangle+\rotate}

            \pgfmathparse{-\midanglerot<90 || -\midanglerot>270? "west" : "east"}
            \edef\textanchor{\pgfmathresult}
            \pgfmathsetmacro\labelshiftdir{1-2*(-\midanglerot>90 && -\midanglerot<270)}

            \fill[\colour] (-\cumnum:\outerradius) arc (-\cumnum:-(\newcumnum):\outerradius) --
            (-\newcumnum:\innerradius) arc (-\newcumnum:-(\cumnum):\innerradius) -- cycle;

            \draw [{Circle[length=2pt]}-,thin] node [append after command={(\midangle:\midradius pt) -- (\midangle:\outerradius + 4ex) -- (\tikzlastnode)}] at (\midangle:\outerradius + 4ex) [xshift=\labelshiftdir*0.2cm,inner sep=0pt, outer sep=0pt, ,anchor=\textanchor]{\small\name: \pgfmathprintnumber{\percentage}\%};

            \global\let\cumnum=\newcumnum
        }

      \end{scope}
    \end{tikzpicture}
}

\wheelchart{
  27.5/BrickRed!75/{\textbf{RISC-V + FPUs}},
  21.8/MidnightBlue!75/{\textbf{Instr. Cache}},
  19.3/Orange!75/{\textbf{RBE}},
  12.3/RoyalBlue!75/{\textbf{TCDM Memory}},
  10.7/PineGreen!75/{\textbf{Interconnect}},
  2.9/Mulberry!75/{\textbf{DMA}},
  2.1/SeaGreen!75/{\textbf{ABB Ctrl}},
  3.4/Gray!75/{\textbf{Other}}%
}








  }
  \caption{Area distribution of \textsc{cluster}.}
  \label{fig:cluster_area}
\end{figure}
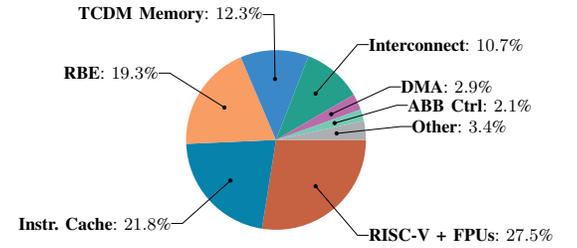

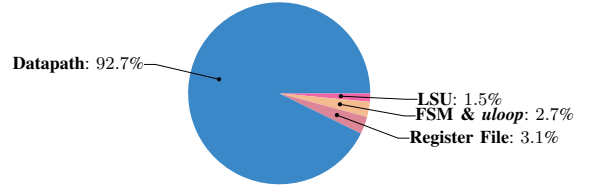
\begin{figure}[tb]
  \centering
  \resizebox{\columnwidth}{!}{%
\def\innerradius{0cm}
\def\innerradiuss{0.5cm}
\def\outerradius{1.6cm}
\def\rotate{0}

\newcommand{\wheelchart}[1]{
    \pgfmathsetmacro{\totalnum}{0}
    \foreach \value/\colour/\name in {#1} {
        \pgfmathparse{\value+\totalnum}
        \global\let\totalnum=\pgfmathresult
    }

    \begin{tikzpicture}

      \pgfmathsetmacro{\wheelwidth}{\outerradius-\innerradiuss}
      \pgfmathsetmacro{\midradius}{(\outerradius+\innerradiuss)/2}

      \begin{scope}[rotate=\rotate]

      \pgfmathsetmacro{\cumnum}{0}
      \foreach \value/\colour/\name in {#1} {
            \pgfmathsetmacro{\newcumnum}{\cumnum + \value/\totalnum*360}

            \pgfmathsetmacro{\percentage}{\value/\totalnum*100}
            \pgfmathsetmacro{\midangle}{-(\cumnum+\newcumnum)/2}
            \pgfmathsetmacro{\midanglerot}{\midangle+\rotate}

            \pgfmathparse{-\midanglerot<90 || -\midanglerot>270? "west" : "east"}
            \edef\textanchor{\pgfmathresult}
            \pgfmathsetmacro\labelshiftdir{1-2*(-\midanglerot>90 && -\midanglerot<270)}

            \fill[\colour] (-\cumnum:\outerradius) arc (-\cumnum:-(\newcumnum):\outerradius) --
            (-\newcumnum:\innerradius) arc (-\newcumnum:-(\cumnum):\innerradius) -- cycle;

            \draw [{Circle[length=2pt]}-,thin] node [append after command={(\midangle:\midradius pt) -- (\midangle:\outerradius + 4ex) -- (\tikzlastnode)}] at (\midangle:\outerradius + 4ex) [xshift=\labelshiftdir*0.2cm,inner sep=0pt, outer sep=0pt, ,anchor=\textanchor]{\small\name: \pgfmathprintnumber{\percentage}\%};

            \global\let\cumnum=\newcumnum
        }

      \end{scope}
    \end{tikzpicture}
}

\wheelchart{
  1.5/Magenta!75/{\textbf{LSU}},
  2.7/cieee2!50/{\textbf{FSM \& \emph{uloop}}},
  3.1/cieee3!50/{\textbf{Register File}},
  92.7/RoyalBlue!75/{\textbf{Datapath}}%
}








  }
  \caption{Post-synthesis area distribution of \rbe{}.}
  \label{fig:rbe_area}
\end{figure}

\begin{figure}[tb]
    \centering \includegraphics[width=0.99\columnwidth]{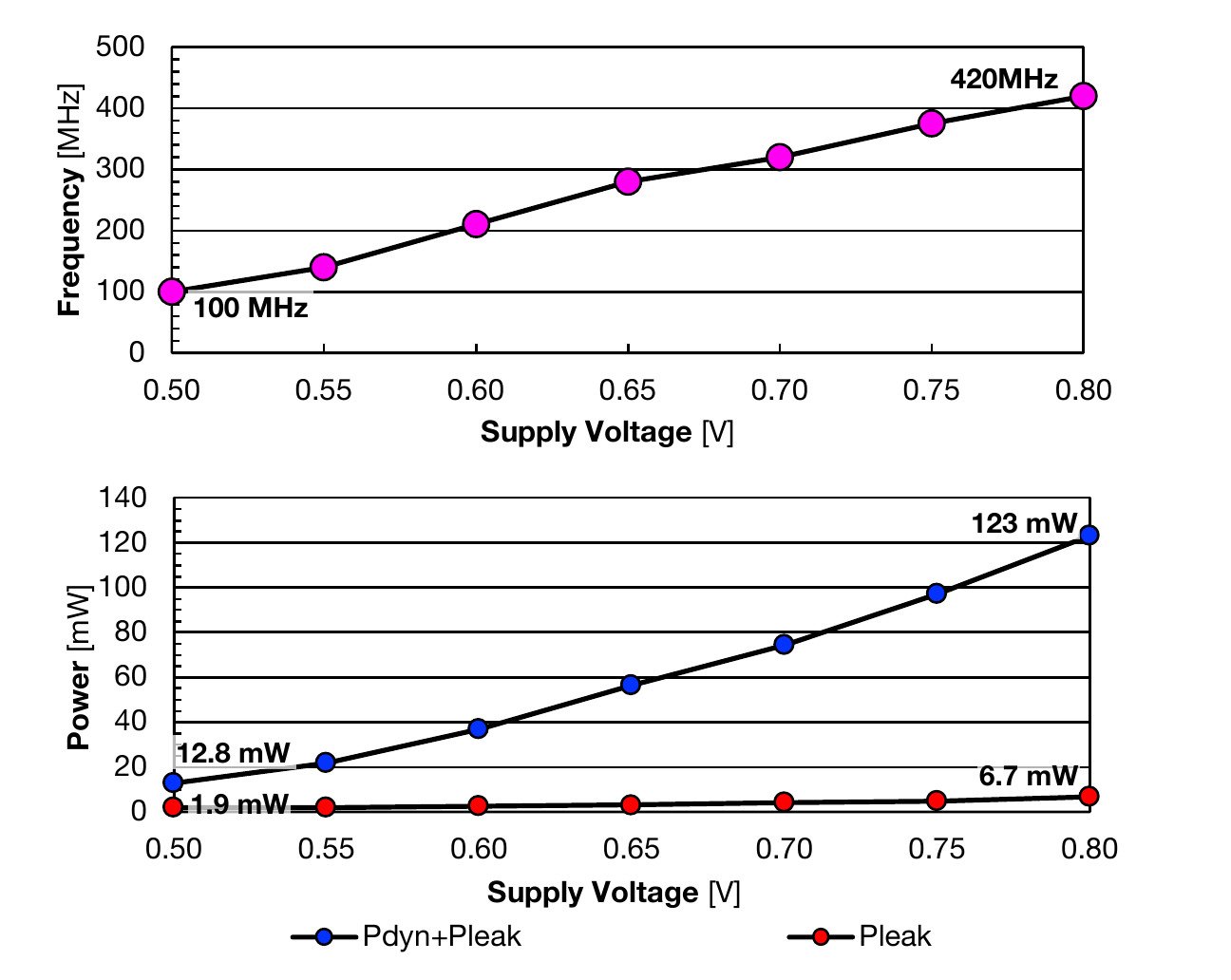}
    \caption{\marsellus{} measured frequency and power sweep while varying $V_{DD}$ (with no ABB).}
    \label{fig:fig4_freq_power}
\end{figure}

The \marsellus{} prototype was fabricated in GlobalFoundries
\SI{22}{\nano\meter} FDX using the flip-well flavor of the technology, which enables \gls{lvt} and \gls{slvt} cells.
A microphotograph of \marsellus{} is shown in Fig.~\ref{fig:fig0_micrograph_recap}.
The prototype is operating with a nominal supply voltage of \SI{0.8}{\volt} with a signoff frequency of \SI{400}{\mega\hertz} for the cluster.
The total die area is \SI{18.7}{\milli\meter^2}, including the full architecture described in Sec.~\ref{sec:architecture} as well as other \glspl{ip} out of the scope of this work.
We focus our analysis and all power measurements on the compute \textsc{cluster}, which occupies an area of \SI{2.42}{\milli\meter^2}, divided as detailed in Fig.~\ref{fig:cluster_area}.
The RISC-V cores, together with the shared instruction cache, occupy almost half of the \textsc{cluster} area, while the \gls{rbe} accelerator takes one fifth. 
For further insight, Fig.~\ref{fig:rbe_area} shows in detail the post-synthesis area breakdown of the \SI{652}{\kilo\gateequi} large \rbe{}; the datapath is with \SI{605}{\kilo\gateequi} (92.7\%) the largest part of \rbe{}.

Fig.~\ref{fig:fig4_freq_power} shows the results of a supply voltage sweep between \SI{0.5}{\volt} and \SI{0.8}{\volt} on the fabricated prototype, without applying the automatic Adaptive Body Biasing described in Section~\ref{sec:abb}.
The maximum frequency achieved at \SI{0.8}{\volt} is \SI{420}{\mega\hertz} (5\% more than the signoff frequency), which scales down to \SI{100}{\mega\hertz} at \SI{0.5}{\volt}.
We characterized the SoC's power consumption in this sweep by using an INT8 matrix-multiplication kernel exploiting the MAC\&LOAD functionality discussed in Section~\ref{sec:macandload}.
At the nominal \SI{0.8}{\volt} supply voltage, this corresponds to a total power consumption of \SI{123}{\milli\watt} (94.6\% dynamic, 5.4\% leakage); dynamic power is reduced by a factor of 10.7$\times$ and leakage by 3.5$\times$ when moving to the lowest-voltage operating point explored (\SI{0.5}{\volt}).

\subsection{Adaptive Body Biasing}

\begin{figure}[tb]%
    \centering
    \includegraphics[width=0.98\columnwidth]{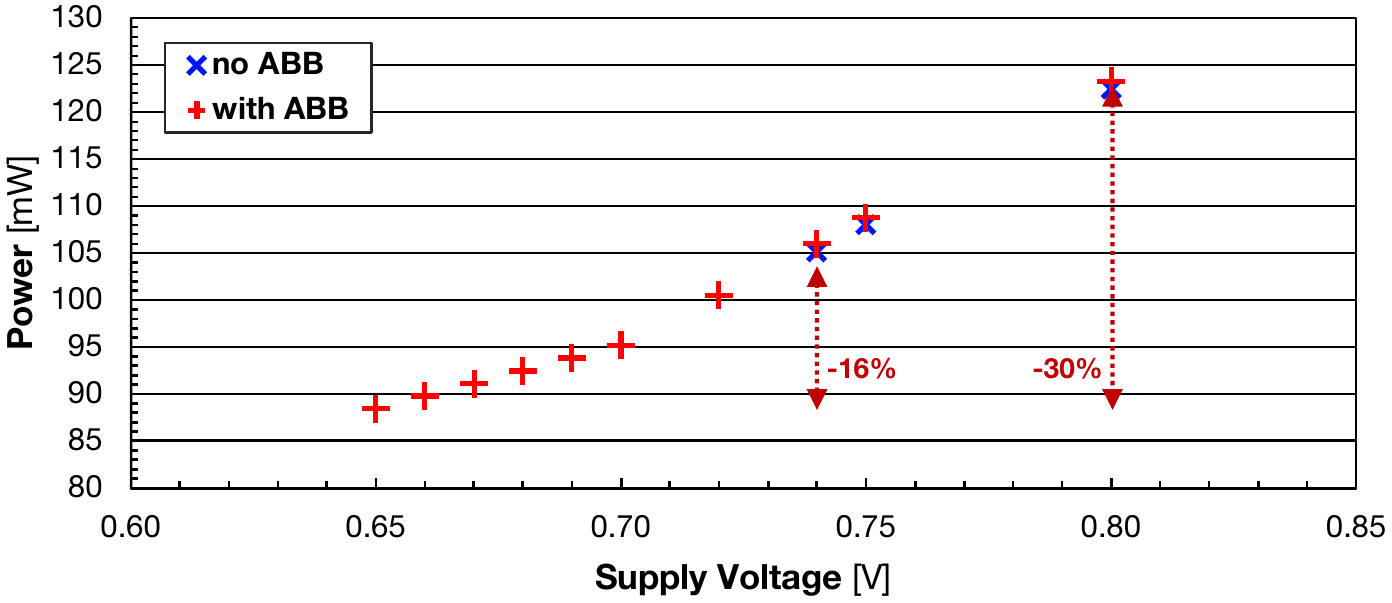}
    \caption{\marsellus{} power measurement with and without applying \gls{abb} for a fixed operating frequency of 400 MHz. Only operating points without timing violations are plotted.}
    \label{fig11a_abb_results}
\end{figure}

\begin{figure}[tb]%
    \centering
    \includegraphics[width=0.98\columnwidth]{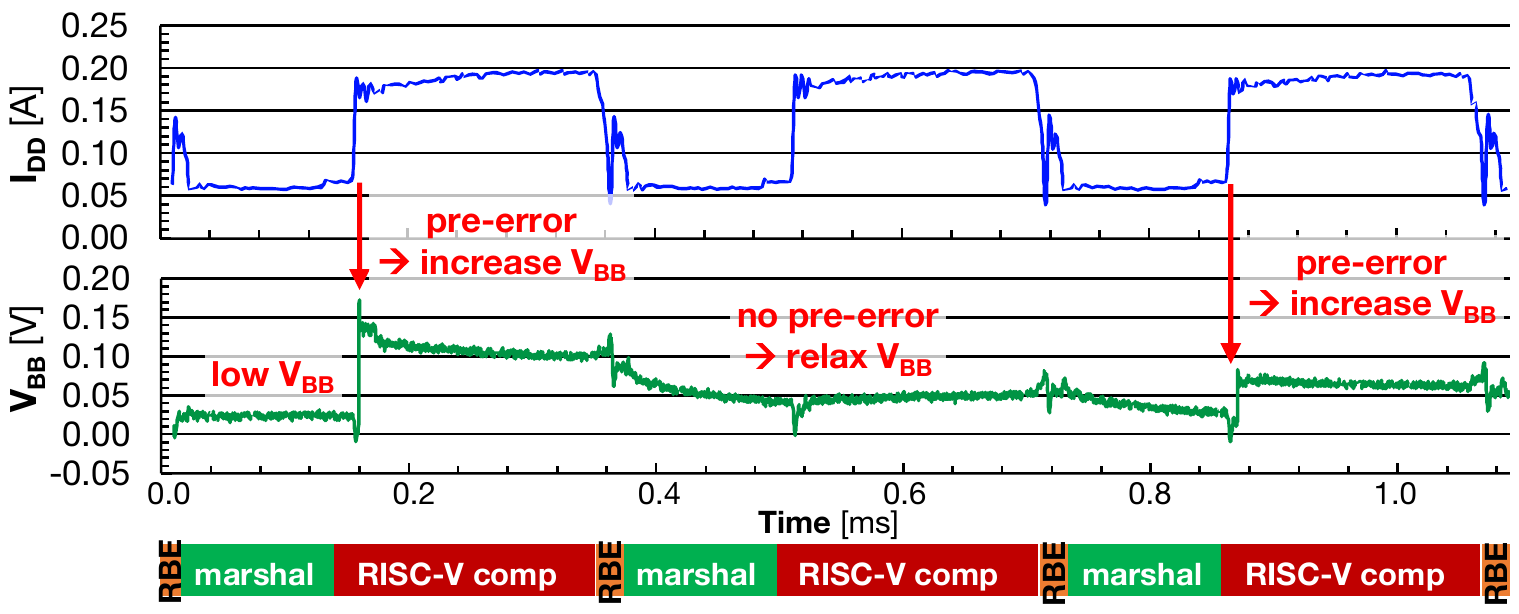}
    \caption{Example of \gls{abb} operation measured on the \marsellus{} prototype, with over-clocking at \SI{470}{\mega\hertz} in the nominal \SI{0.8}{\volt} operating point.}
    \label{fig11b_abb_results}
\end{figure}

\begin{figure}[tb]%
    \centering
    \includegraphics[width=0.98\columnwidth]{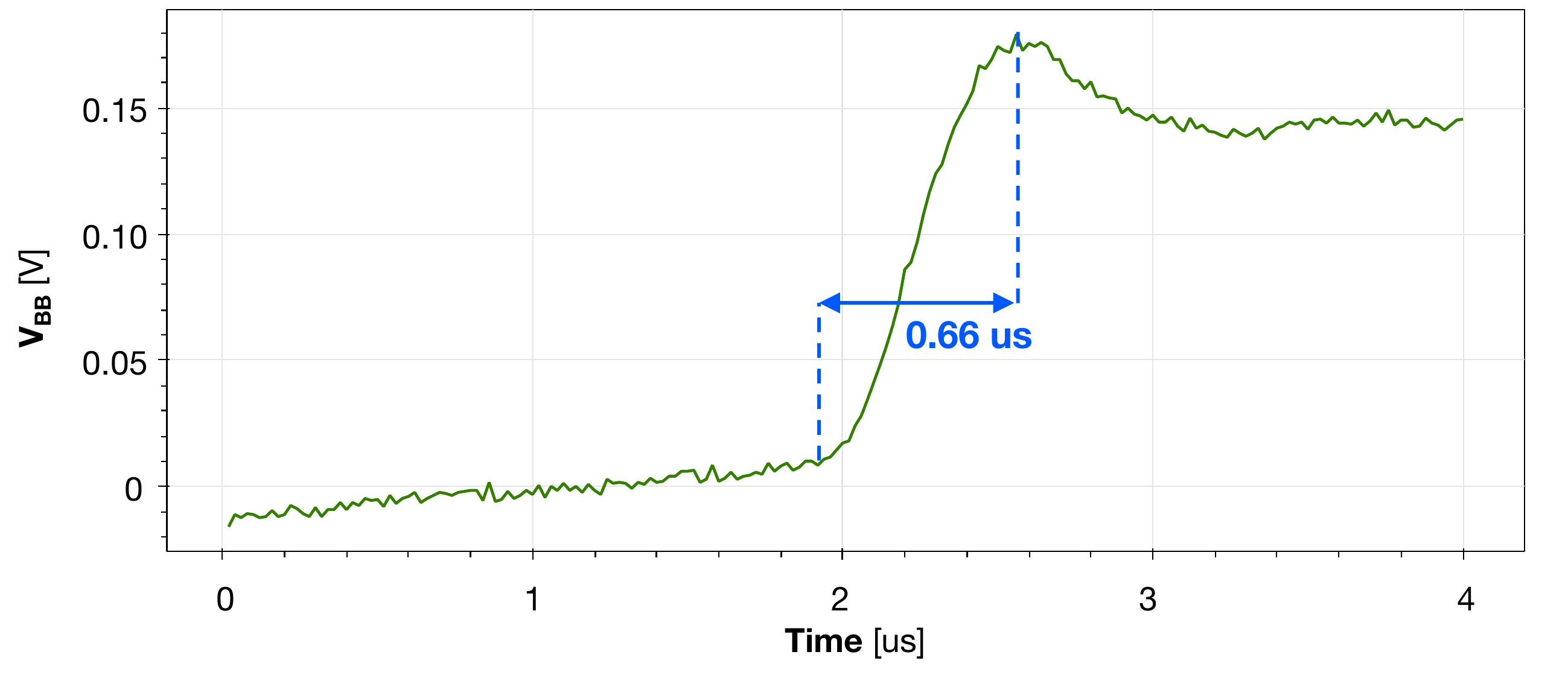}
    \caption{Detail of \SI{4}{\micro\second} of \gls{abb} transition measured on the \marsellus{} prototype, with over-clocking at \SI{470}{\mega\hertz} in the nominal \SI{0.8}{\volt} operating point.}
    \label{fig15_rebuttal_Abb}
\end{figure}

The adoption of flip-well transistors in the \marsellus{} prototype enables using the technique described in Section~\ref{sec:abb} to boost the energy-efficiency of the \marsellus{} \gls{soc} by applying \gls{fbb}.
Fig.~\ref{fig11a_abb_results} shows how \gls{abb} can be exploited to aggressively undervolt the \gls{soc} supply voltage without hitting any performance penalty.
In this experiment, we set the initial operating point at \SI{0.8}{\volt} targeting the signoff frequency of \SI{400}{\mega\hertz}, and then progressively down-scale the voltage in several power measurements performed on a baseline \texttt{Xpulp} INT8 matrix multiplication kernel.
Without applying \gls{abb}, the minimum operating voltage is \SI{0.74}{\volt}, beyond which the \gls{soc} stops working due to timing violations.
\gls{abb} enables to adaptively compensate the increased path delay by detecting pre-errors and correcting them with \gls{fbb}, while retaining the efficiency advantage given by the lower supply voltage.
In this way, it is possible to reduce the supply to \SI{0.65}{\volt} without scaling frequency, with a power reduction of 30\% with respect to the nominal operating point and of 16\% with respect to the \SI{0.74}{\volt} one.

Fig.~\ref{fig11b_abb_results} shows an example of \gls{abb} operation over a synthetic benchmark that alternates three phases of operation: \gls{rbe}-centric and hardware accelerated; low-intensity data marshaling from RISC-V cores; RISC-V based high-intensity computation.
This kind of pattern arises, for example, when \gls{rbe}-supported operators are mixed with non-supported ones, requiring data marshaling to align the \gls{rbe}-centric data layout to a software-centric one.
In the experiment of Fig.~\ref{fig11b_abb_results}, the \marsellus{} \textsc{cluster} was clocked at \SI{470}{\mega\hertz}, a 17.5\% boost with respect to the signoff frequency.
We observe that during the \SI{1}{\milli\second} of operation of the benchmark, the \gls{abb} mechanism is triggered twice to boost \gls{fbb} and enable error-less operation, both times during the higher compute intensity phases.
This is not suprising, as during these phases more near-critical paths are exercised, hence the probability of a pre-error is larger.
Fig.~\ref{fig15_rebuttal_Abb} shows the detail of one such \gls{abb} transition, which has a duration of $\sim$\SI{0.66}{\micro\second} ($\sim$310 clock cycles) after the pre-error is triggered.
When no pre-error is detected in a given time window, the body biasing voltage is progressively relaxed to improve energy efficiency by increasing the effective threshold voltage of flip-well transistors.

\subsection{Performance \& Energy Efficiency}

\subsubsection{RISC-V Performance}
By exploiting \texttt{Xpulpnn}, symmetric 2-bit or 4-bit matrix-vector or matrix-matrix multiplications can be executed in 6$\times$ and $9\times$ less instructions compared to the baseline \texttt{Xpulp}; thanks to the native support for nibble and crumb SIMD operations, \texttt{Xpulpnn} eliminates the overhead to manipulate data to match the lowest precision of the operations available in the ISA of the baseline RI5CY, i.e. 8-bit. MAC\&LOAD accelerated matrix-multiplication kernels further boost the performance by up to 67\%, achieving a DOTP unit utilization as high as 94\%.
As also visible in Fig.~\ref{fig:fig3_xpulpnn} c), after some instructions to initialize the NN-RF, which anyway happens outside the innermost loop, the MAC\&LOAD is able to mask all the explicit loads except one; in combination with data reuse strategy, the innermost loop of the so built kernel outputs 16 accumulators at the cost of a single explicit load operation.
 
Combining the MAC\&LOAD with the \gls{abb} mechanism, \marsellus{} achieves area efficiency on integer linear algebra kernels comparable to that of some \glspl{asic}: 9.63~Gop/s/mm$^2$ in 2\x{}2-bit operation; 4.81~Gop/s/mm$^2$ in 4\x{}4-bit, 2.54~Gop/s/mm$^2$ in 8\x{}8-bit.
At the same time, the RISC-V cores of \marsellus{} retain the full flexibility of floating-point execution when highest precision is needed.
The design is similar to that of Vega~\cite{rossiVegaTenCoreSoC2022}, but the \gls{abb} mechanism and the doubled number of \glspl{fpu} result in 2.1\x{} better absolute performance and slightly better area efficiency (0.37~Gop/s/mm$^2$ against 0.33~Gop/s/mm$^2$ for Vega).

To assess the performance of the RISC-V cluster on non-\gls{ml} applications, we targeted a typical kernel used in \gls{dsp}: \gls{fft}.
\glspl{fft} can constitute a significant fraction of the overall compute workload of an AI-IoT application~\cite{fariselliIntegerOnlyApproximatedMFCC2021}.
We exploited the implementation proposed by Mazzoni~\textit{et~al.}~\cite{mazzoniEfficientTransformAlgorithms2021}, which, measured on a window of 2048 samples, achieves a peak throughput of 4.69 FLOp/cycle when parallelized on 16 cores, leading to a peak performance of 1.97 GFLOPS in the \SI{0.8}{\volt},\SI{420}{\mega\hertz} operating point, and a top efficiency 36 GFLOPS/W in the \SI{0.5}{\volt},\SI{100}{\mega\hertz} one.

\begin{figure}[tb]
    \centering \includegraphics[width=0.99\columnwidth]{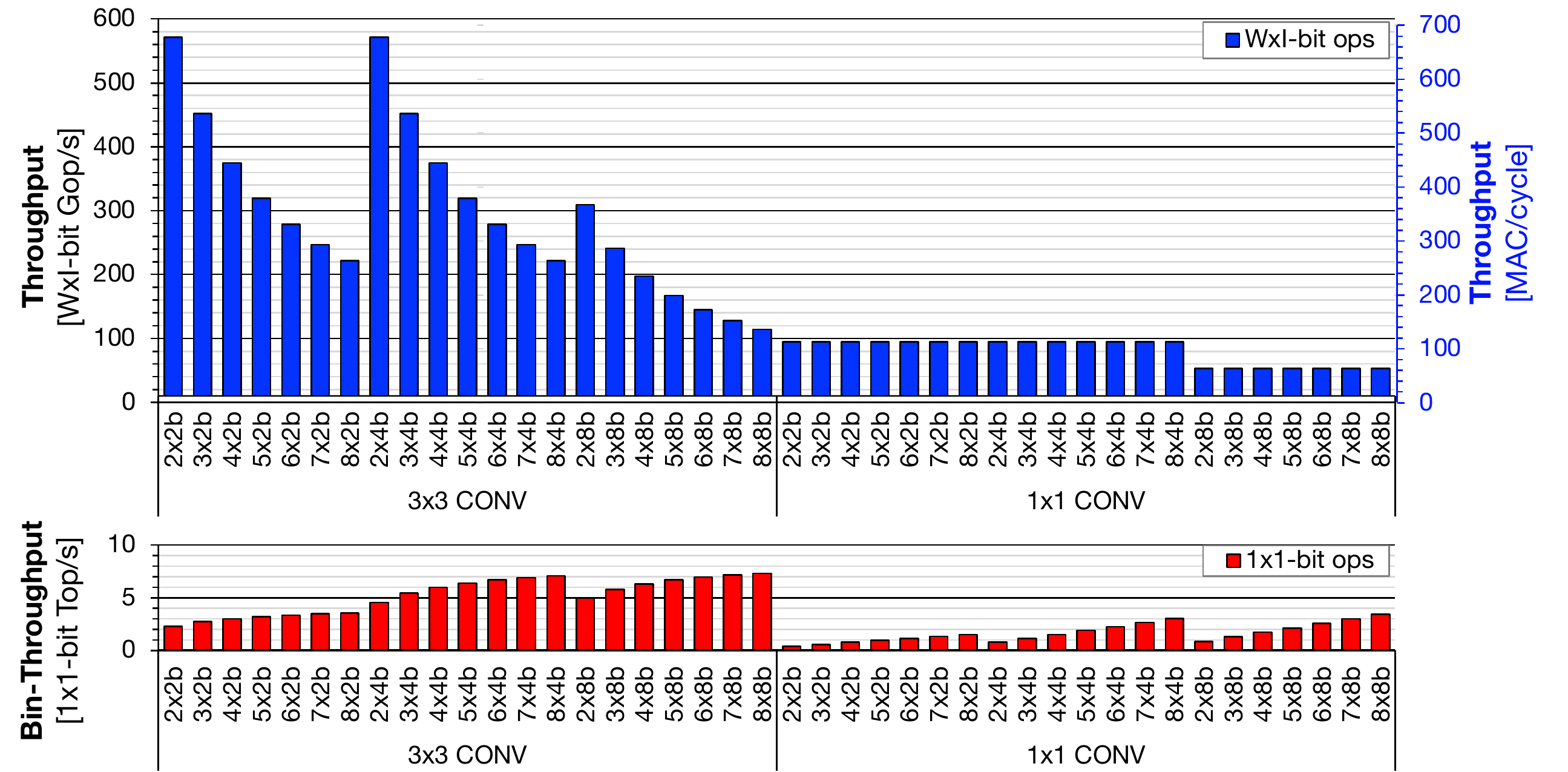}
    \caption{Main \texttt{\bfseries\color{Orange}LOAD}-\texttt{\bfseries\color{RoyalBlue}COMPUTE} loop throughput for 3\x{}3 and 1\x{}1 convolutions in terms of \textit{W}\x{}\textit{I}-bit~(blue) and 1\x{}1-bit~(red) operations running on \rbe{} at 0.8V and a nominal frequency of 420 MHz, computing a layer with $K_{in}=64$, $K_{out}=64$, $H=W=3$. The right axis reports performance scaled in \textit{W}\x{}\textit{I}-bit MAC/cycle.}
    \label{fig:fig6_rbe_detail}
\end{figure}
\subsubsection{RBE Performance}
\label{sec:rbe_perf}

As detailed in Section~\ref{sec:rbe}, the \gls{rbe} can be used in several different configurations in terms of operation and precision, which correspond to different throughput.
Fig.~\ref{fig:fig6_rbe_detail} analyses the performance of \gls{rbe} in many supported configurations\footnote{Configurations with non-power-of-two \textit{I} are omitted due to space reasons, but follow the same trends highlighted in Fig.~\ref{fig:fig6_rbe_detail} and in the discussion.} when computing a convolutional layer with 64 output channels ($K_{out}$), 64 input channels ($K_{in}$), and output spatial size 3\x{}3.
We focus on two different metrics: actual throughput at the  \textit{W}\x{}\textit{I}-bit precision, which is the target application-aware performance metric, and ``raw'' throughput at 1\x{}1-bit, which instead measures the raw utilization of the compute resources over the full computation loop (considering both the \texttt{\bfseries\color{Orange}LOAD} and \texttt{\bfseries\color{RoyalBlue}COMPUTE} phases).

Several effects are visible.
First of all, the 1\x{}1-bit throughput is higher when activations are 4-bit or larger, as in that case the \rbe{} all BinConvs in a Block are utilized.
Configuration with \textit{I}=8-bit, however, result in a $\sim$50\% actual throughput reduction because their contributions are split in consecutive iterations.
3\x{}3 convolutions suffer from little overhead introduced by the \texttt{\bfseries\color{Orange}LOAD} phase, while 1\x{}1 convolutions are hit more heavily, due to the fact that the duration of their \texttt{\bfseries\color{RoyalBlue}COMPUTE} phase is much shorter and comparable with \texttt{\bfseries\color{Orange}LOAD}.
Finally, in 3\x{}3 mode actual throughput gets higher when reducing the \textit{W} size (although the binary throughput is decreased): this is due to the fact that the weight-bit dimension is serialized in this mode.
Changing \textit{W} does not impact performance in 1\x{}1 convolutions because this dimension is parallelized across the Blocks in each \rbe{} Core and hence only affects the Core's utilization.
Overall, the highest 1\x{}1-bit throughput achieved is $\sim$7100~1\x{}1-bit~Top/s in the \textit{W}=8, \textit{I}=4 configuration of the 3\x{}3 convolution mode.
The highest throughput, on the other hand, is of 571~Gop/s in the \textit{W}=2, \textit{I}=4 configuration of 3\x{}3 convolution.

\subsubsection{RISC-V \& RBE Performance/Efficiency}
\begin{figure}
    \centering \includegraphics[width=0.95\columnwidth]{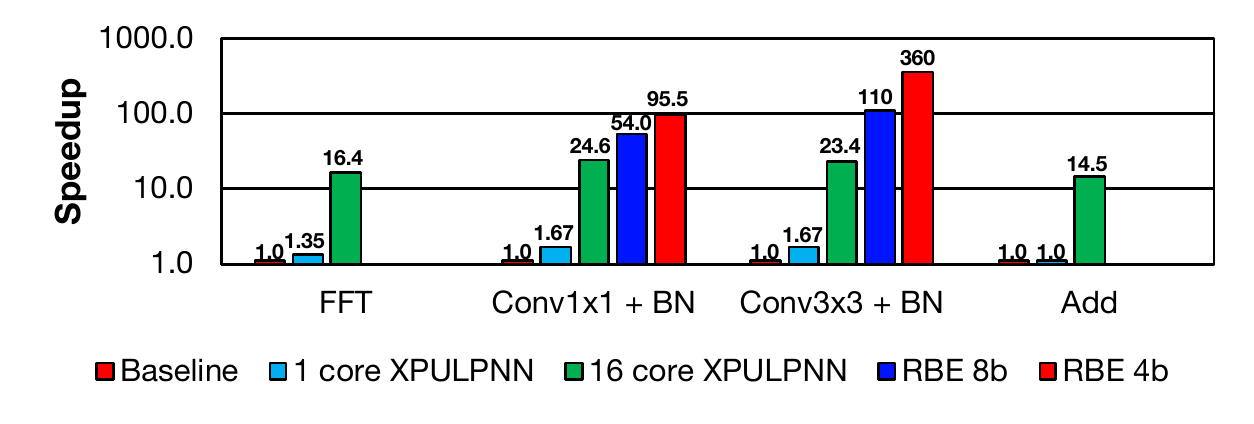}
    \caption{Speedup of AI and non-AI tasks on \marsellus{} \textsc{cluster} vs execution on \marsellus{} \textsc{soc}.}
    \label{fig:fig16_speedup}
\end{figure}

\begin{figure}
    \centering \includegraphics[width=0.95\columnwidth]{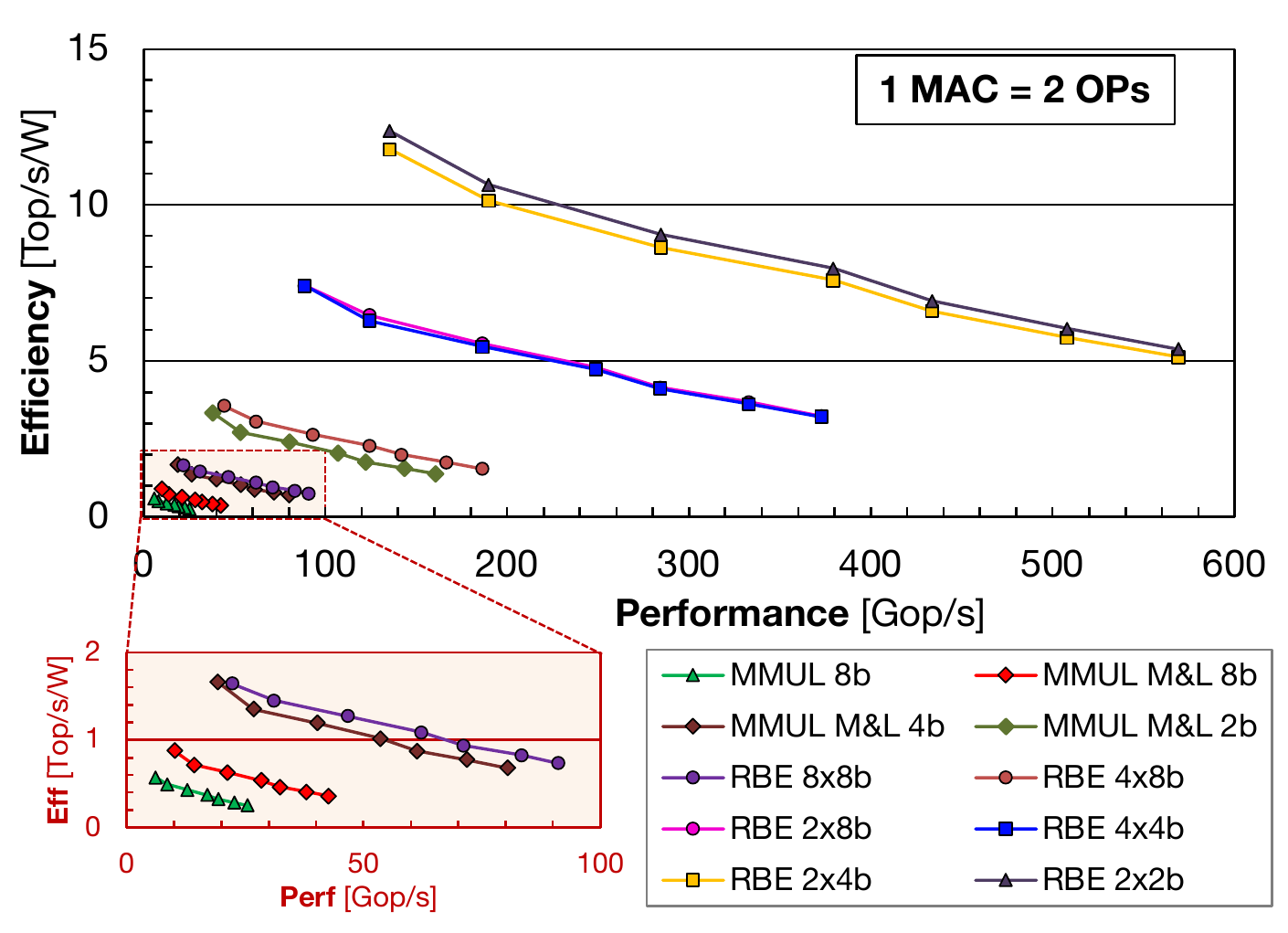}
    \caption{Energy efficiency versus performance for 3\x{}3 convolutions on RBE and MMUL on RISC-V cores measured on \marsellus{}.}
    \label{fig:fig5_perf_eff}
\end{figure}

Fig.~\ref{fig:fig16_speedup} collects overall speedups for four tasks: floating-point 2048-point FFT, 8-bit 1$\times$1 pointwise and 3$\times$3 convolutional layers including batch-normalization on a $9\times 9\times 64$ output space with 64 input channels, tensor addition of two $9\times 9\times 64$ tensors.
Results are shown for the baseline \texttt{XpulpV2} core in the \textsc{soc}, for a single \textsc{cluster} core, for the full \textsc{cluster} and, where possible, for execution with \gls{rbe} (8- and 4-bit).
We note that this speedup can be reduced in pathological conditions; for example, a Conv1$\times$1 on a single input channel will be $\sim$2$\times$ faster on the RISC-V cluster than on RBE.

Fig.~\ref{fig:fig5_perf_eff} gives an overview of the \textsc{cluster}-level measured energy efficiency versus achieved performance without applying \gls{abb}, while sweeping frequency and supply voltage as indicated in Fig.~\ref{fig:fig4_freq_power}.
Each curve reflects measurements at different operating points on one of three different benchmarks: parallel \riscv{} matrix multiplication (MMUL); parallel \riscv{} MMUL exploiting the \texttt{Xpulpnn} extensions (MMUL M\&L); and RBE-based 3\x{}3 convolution kernels.
All measurements include the full \textsc{cluster} power.
The baseline MMUL kernel achieves a performance of 25.45~Gop/s and efficiency of 250~Gop/s/W in the nominal operating point (\SI{0.8}{\volt}), which scales to 6.06~Gop/s and 580~Gop/s/W, respectively, when downascaling $V_{DD}$ to \SI{0.5}{\volt}.
The architectural improvements introduced with \texttt{Xpulpnn} and the MAC\&LOAD mechanism improve performance and efficiency by 67\% and 51\%, respectively.
Introducing aggressive quantization at 4-bit and 2-bit leads to more savings: 3.2\x{} and 2.9\x{} in performance and efficiency, respectively, comparing the MAC\&LOAD 4-bit versus the MMUL 8-bit baseline; 6.3\x{} and 5.7\x{}, respectively, when considering the 2-bit MAC\&LOAD version.
%

The highest-precision (8\x{}8-bit) \gls{rbe} configuration has a throughput of 91~Gop/s and an efficiency of 740~Gop/s/W in the nominal operating point.
Voltage downscaling boosts efficiency up to 1.64~Top/s/W at the expense of a significant loss of performance (down  to 22~Gop/s).
In the \rbe{} case, aggressive quantization can be used with considerable freedom; Fig.~\ref{fig:fig5_perf_eff} restricts analysis to power-of-two configurations where \textit{W}$\leq$\textit{I}, which are the most common in quantized CNNs.
When scaling \textit{W} and \textit{I} to the minimum of 2 bits, performance and efficiency are maximized, yielding 569~Gop/s and 5.37~Top/s/W, respectively, in the \SI{0.8}{\volt} operating point and 136~Gop/s and 12.36~Top/s/W, respectively, in the \SI{0.5}{\volt} operating point.



\begin{figure}[tb]
    \centering\includegraphics[width=0.92\columnwidth]{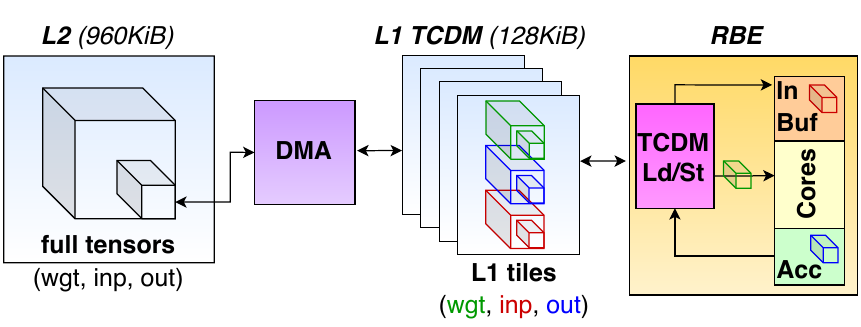}
    \caption{DNN execution with data tiling on \marsellus{}.}
    \label{fig:fig12_dnn_tiling}
\end{figure}

\begin{figure}[tb]
    \centering\includegraphics[width=0.98\columnwidth]{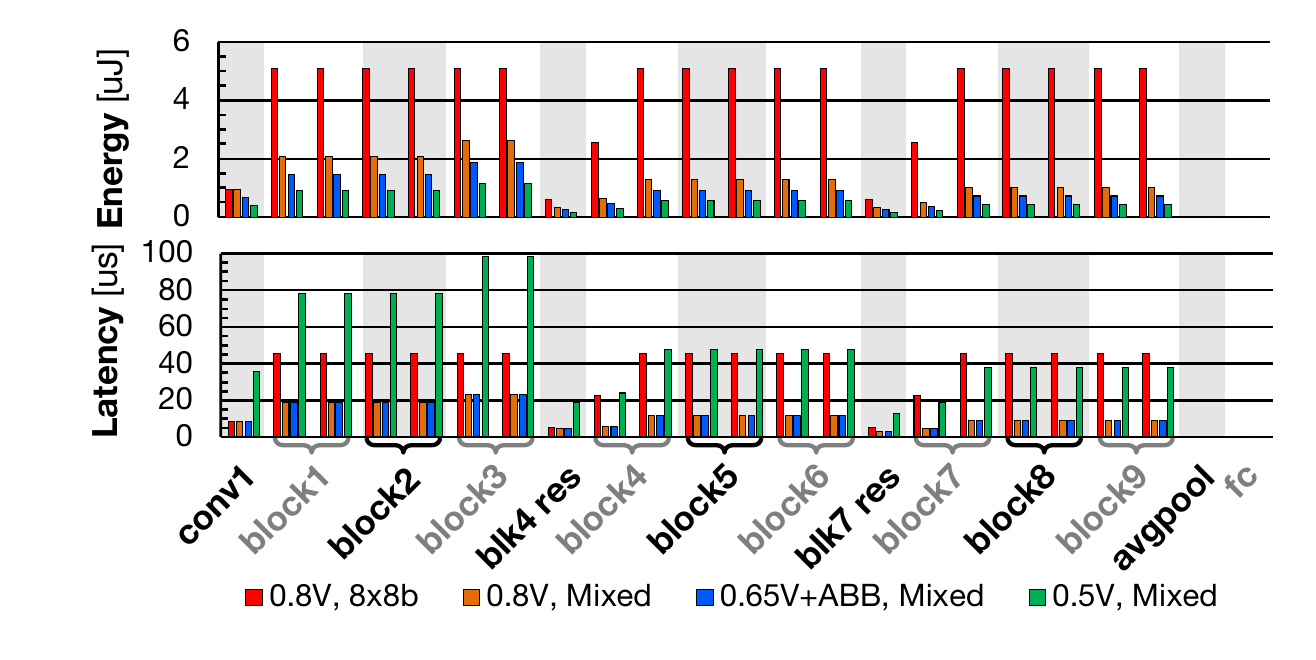}
    \caption{Layer-wise breakdown of latency and energy consumption of an end-to-end ResNet-20 network on the CIFAR-10 dataset running on \marsellus{} with 8-bit and mixed precision quantization for different operating points.}
    \label{fig:fig12_resnet20_detail}
\end{figure}

\begin{figure}[tb]
    \centering\includegraphics[width=0.98\columnwidth]{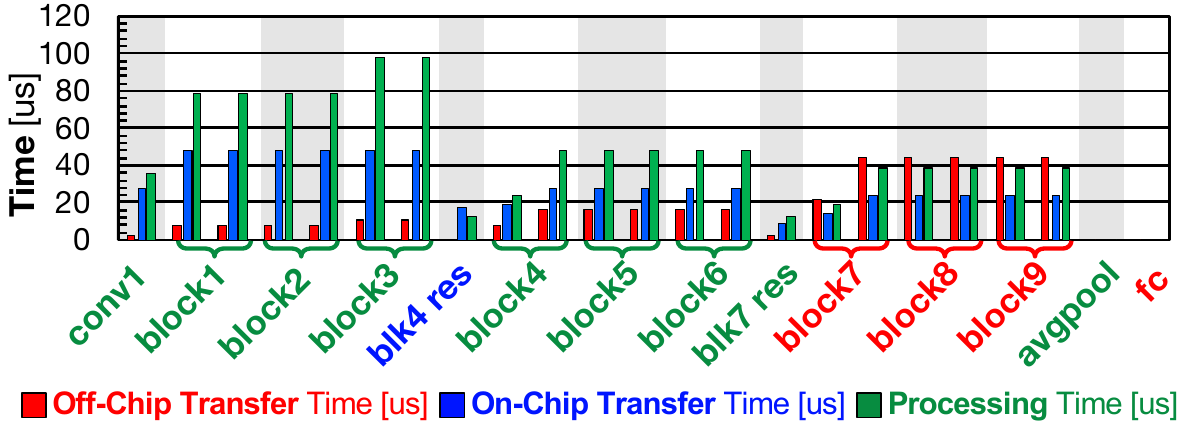}
    \caption{Detail of ResNet-20/CIFAR-10 in the \SI{0.5}{\volt} mixed precision configuration, showing latency of off-chip and on-chip transfers and processing (compute + tiling overheads). Latencies are fully overlapped and thus the tallest bar in each group defines the latency of a layer. Red/blue/green-labeled groups of layers are off-chip/on-chip/compute dominated, respectively.}
    \label{fig:fig14_rebuttal}
\end{figure}
\section{Mapping DNNs on Marsellus}
\label{sec:dnn}

To deploy \glspl{dnn} on \marsellus{}, 
we exploit a modified version of the QuantLab open-source library\footnote{\url{https://github.com/pulp-platform/quantlab}} built on PyTorch for quantization.
QuantLab exports a fully quantized \gls{onnx} graph which can then be mapped on \marsellus{} by the back-end DORY~\cite{burrelloDORYAutomaticEndtoEnd2021} tool.
\gls{dnn} tensors need to be tiled between the various levels of the memory hierarchy, which are explicitly managed~\cite{burrelloDORYAutomaticEndtoEnd2021}, implementing the mechanism shown in Fig.~\ref{fig:fig12_dnn_tiling}.
With data tiling, the execution of the full layers is split in smaller chunks that can fit the available L1 \gls{tcdm} memory budget (\SI{128}{\kibi\byte}).
The \textsc{cluster} \gls{dma} loads each weight and input activation tile from the L2 memory into the L1 \gls{tcdm} memory where the RISC-V cores and \rbe{} can work on them.
Similarly, the final output activations are stored by the \gls{dma} from the L1 into the L2 memory.
A double-buffering mechanism is used to perform \gls{dma} transfers autonomously, while the cores and/or \rbe{} are computing.

We deployed layer-by-layer an end-to-end ResNet-20 trained on the CIFAR-10 dataset on \marsellus{}, focusing on exploiting the \rbe{} accelerator.
Off-chip memory accesses are modeled using an analytical model of I/O obtained from data of a previous prototype using the same technology node~\cite{rossiVegaTenCoreSoC2022}.
Exploiting the HAWK~\cite{dong2019hawq} quantization scheme, it is possible to quantize the network weights at arbitrary precisions (with 2-, 3-, 6-, 8-bits) and activations at 4- and 8-bit activations; 
as reported by Dong~\textit{et~al.}, this quantization-aware training scheme leads to a negligible accuracy loss (from 92.4\% to 92.2\%).
Exploiting the bit-flexible \gls{dnn} support from \rbe{}, this aggressive quantization scheme yields significant energy efficiency improvement, saving 68\% of the execution energy, down to $\sim$\SI{28}{\micro\joule}.

\begin{table*}[tb]
\centering
\caption{\gls{abb} methods in the \gls{soa}.}
\label{tab:abb}
\begin{adjustbox}{width=\textwidth,totalheight=\textheight,keepaspectratio}
\begin{tabular}{llllll}
\toprule
 &
  \textbf{Technology Node} &
  \textbf{Prototype} &
  \textbf{Area} &
  \textbf{Best power gain} &
  \textbf{Automatic tuning method} \\ \midrule
\textit{Moursy et   al.~\cite{moursy35021mm2PVTAware2021}} &
  22nm FDX &
  Cortex-M4F (core+memory) &
  2 mm &
  -19.9\% &
  On-Chip Monitors + ABB-generator \\
\textit{Rossi et al.~\cite{rossi60GOPS8V2016}} &
  28nm FD-SOI &
  4-core PULP cluster &
  3 mm$^2$ &
  -43\% (sleep) &
  None \\
\textit{SleepRunner~\cite{bolSleepRunner28nmFDSOI2021}} & 28nm FD-SOI & Cortex-M0 MCU & 0.6 mm$^2$ & - & Unified Frequency/Bias Regulators \\
\textit{Akgul et   al.~\cite{akgulPowerManagementDVFS2014}} &
  28nm FD-SOI &
  32-bit VLIW DSP &
  - &
  -17\% &
  Offline software \\
\textit{Quelen et   al.~\cite{quelen5mW0067mm2Automatic2018}} &
  28nm FD-SOI &
  0.1-2mm2 digital core &
  2 mm$^2$ &
  -32\% &
  On-Chip Monitors + ABB-generator \\
\textit{Marsellus (our work)} &
  22nm FDX &
  17 RISC-V + RBE &
  2.42 mm$^2$ &
  -30\% &
  On-Chip Monitors + ABB-generator \\
  \bottomrule
\end{tabular}%
\end{adjustbox}
\end{table*}
%
\begin{table*}[tb]
  \centering
  \caption{Comparison of \marsellus{} with related work.}
  \begin{adjustbox}{width=\textwidth,totalheight=\textheight,keepaspectratio}
      \begin{threeparttable}
        \begin{tabular}{@{}llllll@{}}
          \toprule
            & Vega~\cite{rossiVegaTenCoreSoC2022} & SAMURAI~\cite{miro-panadesSamurAIVersatileIoT2022} & DIANA~\cite{houshmandDIANAEndtoEndHybrid2023} & QNAP~\cite{mo12TOPSQuantized2022} & \textit{\marsellus{}} (this work) \\
          \midrule
          \textit{Technology}          & 22nm FDX  & 28nm FD-SOI & 22nm FDX+AIMC & 28nm      &  22nm FDX               \\
          \textit{Die Area}            & 10mm$^2$  & 4.5mm$^2$   & 10.24mm$^2$         & 1.9mm$^2$ &  18.7mm$^2$ (2.42mm$^2$) \tnote{a} \\
          \textit{Applications}        & \gls{iot} GP+DNN & \gls{iot} GP+DNN & \acrshort{ai-iot} & DNN ASIC& \gls{iot} GP+DNN + \acrshort{ai-iot} \\
          \multirow{1}{*}{\textit{SRAM}}
                & \multirow{1}{*}{\begin{tabular}[c]{@{}l@{}} 128KiB L1 + 1.6MiB L2 \end{tabular}}
                & \multirow{1}{*}{\begin{tabular}[c]{@{}l@{}} 464KiB \end{tabular}}
                & \multirow{1}{*}{\begin{tabular}[c]{@{}l@{}} 896KiB \end{tabular}}
                & \multirow{1}{*}{\begin{tabular}[c]{@{}l@{}} 206KiB \end{tabular}}
                & \multirow{1}{*}{\begin{tabular}[c]{@{}l@{}} 128KiB L1 + 1MiB L2 \end{tabular}} \\
          \midrule
          \multirow{3}{*}{\textit{Cores}} & \multirow{3}{*}{\begin{tabular}[c]{@{}l@{}} 10\x{}RV32IMCFXpulp \\ \quad + HWCE \\ \end{tabular}}
                                & \multirow{3}{*}{\begin{tabular}[c]{@{}l@{}} 1\x{}RV32IMCFXpulp \\ \quad + digital Accel \\ \end{tabular}}
                                & \multirow{3}{*}{\begin{tabular}[c]{@{}l@{}} 1\x{}RV32IMCFXpulp \\ \quad + digital Accel \\ \quad + AIMC \end{tabular}}
                                & \multirow{3}{*}{\begin{tabular}[c]{@{}l@{}} digital Accel \end{tabular}}
                                & \multirow{3}{*}{\begin{tabular}[c]{@{}l@{}} 16\x{}RV32IMCFXpulpnn \\ \quad + 1\x{} RV32IMCFXpulp \\ \quad + RBE \end{tabular}} \\ \\ \\
          \midrule
          \multirow{1}{*}{\textit{INT precisions}} & \multirow{1}{*}{\begin{tabular}[c]{@{}l@{}} 8,16,32 \end{tabular}}
                                          & \multirow{1}{*}{\begin{tabular}[c]{@{}l@{}} 8,16,32 \end{tabular}}
                                          & \multirow{1}{*}{\begin{tabular}[c]{@{}l@{}} 2,4,8,16,32 \end{tabular}}
                                          & \multirow{1}{*}{\begin{tabular}[c]{@{}l@{}} 8 \end{tabular}}
                                          & \multirow{1}{*}{\begin{tabular}[c]{@{}l@{}} 2,4,8,16,32, RBE: 2-8 \end{tabular}} \\
          \textit{FP precisions}       & FP32, FP16, BF16  & -         & -         & -         & FP32, FP16, BF16      \\
          \midrule
          \textit{Supply Voltage}      & 0.5-0.8V          & 0.45-0.9V & 0.5-0.9V  & 0.6-0.9V  & 0.5-0.8V              \\
          \textit{Max. Frequency}      & 450MHz            & 350MHz    & 320MHz    & 470MHz    & 420MHz                \\
          \textit{Power Range}         & 1.7$\mu$W-49.4mW  & 6.4$\mu$W-96mW  & 10mW-129mW & 19.4-131mW & 12.8mW-123mW  \\
          \midrule
          \multirow{2}{*}{\begin{tabular}[c]{@{}l@{}} \textit{Best SW (INT) Perf} \end{tabular}}
                & \multirow{2}{*}{\begin{tabular}[c]{@{}l@{}} 15.6 Gop/s \tnote{b} \end{tabular}}
                & \multirow{2}{*}{\begin{tabular}[c]{@{}l@{}}  1.5 Gop/s \tnote{c} \end{tabular}}
                & \multirow{2}{*}{\begin{tabular}[c]{@{}l@{}} - \end{tabular}}
                & \multirow{2}{*}{\begin{tabular}[c]{@{}l@{}} - \end{tabular}}
                & \multirow{2}{*}{\begin{tabular}[c]{@{}l@{}} \textbf{180  Gop/s \tnote{d}} \\ \quad (2\x{}2b, 0.8V+ABB)\end{tabular}} \\ \\
                \midrule
          \multirow{2}{*}{\begin{tabular}[c]{@{}l@{}} \textit{Best SW (INT) Area Eff} \end{tabular}}
                & \multirow{2}{*}{\begin{tabular}[c]{@{}l@{}} 1.56 Gop/s/mm$^2$ \tnote{b} \end{tabular}}
                & \multirow{2}{*}{\begin{tabular}[c]{@{}l@{}}  0.33 Gop/s/mm$^2$ \tnote{c} \end{tabular}}
                & \multirow{2}{*}{\begin{tabular}[c]{@{}l@{}} - \end{tabular}}
                & \multirow{2}{*}{\begin{tabular}[c]{@{}l@{}} - \end{tabular}}
                & \multirow{2}{*}{\begin{tabular}[c]{@{}l@{}} \textbf{9.63 Gop/s/mm$^2$ \tnote{d}} \\ \quad (2\x{}2b, 0.8V+ABB)\end{tabular}} \\ \\
                \midrule
          \multirow{2}{*}{\begin{tabular}[c]{@{}l@{}} \textit{Best SW (INT) Energy Eff} \end{tabular}}
                & \multirow{2}{*}{\begin{tabular}[c]{@{}l@{}} 614 Gop/s/W \\ \quad @ 7.6 Gop/s \tnote{b} \end{tabular}}
                & \multirow{2}{*}{\begin{tabular}[c]{@{}l@{}} 230 Gop/s/W \\ \quad @ 110 MOp/s \tnote{c} \end{tabular}}
                & \multirow{2}{*}{\begin{tabular}[c]{@{}l@{}} - \end{tabular}}
                & \multirow{2}{*}{\begin{tabular}[c]{@{}l@{}} - \end{tabular}}
                & \multirow{2}{*}{\begin{tabular}[c]{@{}l@{}} \textbf{3.32 Top/s/W} \\ \quad @ 19 Gop/s \mtnote{d} (2\x{}2b, 0.5V)\end{tabular}} \\ \\
          \midrule
          \multirow{1}{*}{\begin{tabular}[c]{@{}l@{}} \textit{Best SW (FP16) Perf} \end{tabular}}
                & \multirow{1}{*}{\begin{tabular}[c]{@{}l@{}} 3.3 Gflop/s \tnote{b} \end{tabular}}
                & \multirow{1}{*}{\begin{tabular}[c]{@{}l@{}} - \end{tabular}}
                & \multirow{1}{*}{\begin{tabular}[c]{@{}l@{}} - \end{tabular}}
                & \multirow{1}{*}{\begin{tabular}[c]{@{}l@{}} - \end{tabular}}
                & \multirow{1}{*}{\begin{tabular}[c]{@{}l@{}} \textbf{6.9 Gflop/s} \mtnote{d} (0.8V+ABB) \end{tabular}}  \\
          \midrule
          \multirow{1}{*}{\begin{tabular}[c]{@{}l@{}} \textit{Best SW (FP16) Area Eff} \end{tabular}}
                & \multirow{1}{*}{\begin{tabular}[c]{@{}l@{}} 0.33 Gflop/s/mm$^2$ \tnote{b} \end{tabular}}
                & \multirow{1}{*}{\begin{tabular}[c]{@{}l@{}} - \end{tabular}}
                & \multirow{1}{*}{\begin{tabular}[c]{@{}l@{}} - \end{tabular}}
                & \multirow{1}{*}{\begin{tabular}[c]{@{}l@{}} - \end{tabular}}
                & \multirow{1}{*}{\begin{tabular}[c]{@{}l@{}} \textbf{0.37 Gflop/s/mm$^2$} \mtnote{d} (0.8V+ABB) \end{tabular}}  \\
          \midrule
          \multirow{2}{*}{\begin{tabular}[c]{@{}l@{}} \textit{Best SW (FP16) Energy Eff} \end{tabular}}
                & \multirow{2}{*}{\begin{tabular}[c]{@{}l@{}} 129 Gflop/s/W \\ \quad @ 1.7 Gflop/s \tnote{b} \end{tabular}}
                & \multirow{2}{*}{\begin{tabular}[c]{@{}l@{}} - \end{tabular}}
                & \multirow{2}{*}{\begin{tabular}[c]{@{}l@{}} - \end{tabular}}
                & \multirow{2}{*}{\begin{tabular}[c]{@{}l@{}} - \end{tabular}}
                & \multirow{2}{*}{\begin{tabular}[c]{@{}l@{}} \textbf{207 Gflop/s/W} \\ \quad @ 3.1 Gflop/s \tnote{d} \end{tabular}} \\ \\
          \midrule
          \multirow{2}{*}{\begin{tabular}[c]{@{}l@{}} \textit{Best HW-Accel Perf} \end{tabular}}
                & \multirow{2}{*}{\begin{tabular}[c]{@{}l@{}} 32.2 Gop/s \end{tabular}}
                & \multirow{2}{*}{\begin{tabular}[c]{@{}l@{}} 36.0 Gop/s \end{tabular}}
                & \multirow{2}{*}{\begin{tabular}[c]{@{}l@{}} digital: 180 Gop/s  \\ AIMC: \textbf{29.5 Top/s}\end{tabular}}
                & \multirow{2}{*}{\begin{tabular}[c]{@{}l@{}} 140 Gop/s \end{tabular}}
                & \multirow{2}{*}{\begin{tabular}[c]{@{}l@{}} \textbf{637 Gop/s} \\ \quad (2\x{}2b, 0.8V+ABB)\end{tabular}} \\ \\
          \midrule
          \multirow{2}{*}{\begin{tabular}[c]{@{}l@{}} \textit{Best HW-Accel Area Eff} \end{tabular}}
                & \multirow{2}{*}{\begin{tabular}[c]{@{}l@{}} 3.22 Gop/s/mm$^2$ \end{tabular}}
                & \multirow{2}{*}{\begin{tabular}[c]{@{}l@{}} 8.0 Gop/s/mm$^2$ \end{tabular}}
                & \multirow{2}{*}{\begin{tabular}[c]{@{}l@{}} digital: 17.6 Gop/s/mm$^2$  \\ AIMC: \textbf{2.9 Top/s/mm$^2$}\end{tabular}}
                & \multirow{2}{*}{\begin{tabular}[c]{@{}l@{}} \textbf{73.7 Gop/s/mm$^2$} \end{tabular}}
                & \multirow{2}{*}{\begin{tabular}[c]{@{}l@{}} 34.1 Gop/s/mm$^2$ \\ \quad (2\x{}2b, 0.8V+ABB)\end{tabular}} \\ \\
          \midrule
          \multirow{2}{*}{\begin{tabular}[c]{@{}l@{}} \textit{Best HW-Accel Energy Eff} \end{tabular}}
              & \multirow{2}{*}{\begin{tabular}[c]{@{}l@{}} 1.3 Top/s/W \\ \quad @ 15.6 Gop/s \end{tabular}}
              & \multirow{2}{*}{\begin{tabular}[c]{@{}l@{}} 1.3 Top/s/W \\ \quad @ 2.8 Gop/s \end{tabular}}
              & \multirow{2}{*}{\begin{tabular}[c]{@{}l@{}} digital: 4.1 Top/s/W \\ AIMC: \textbf{600 Top/s/W} \end{tabular}}
              & \multirow{2}{*}{\begin{tabular}[c]{@{}l@{}} \textbf{12.6 Top/s/W} \\ \quad @ 140 Gop/s (8b) \end{tabular}}
              & \multirow{2}{*}{\begin{tabular}[c]{@{}l@{}} \textbf{12.4 Top/s/W} \\ \quad @ 136 Gop/s (2\x{}2b, 0.5V)\end{tabular}} \\ \\
          \midrule
          \multirow{1}{*}{\begin{tabular}[c]{@{}l@{}} \textit{ResNet-20/CIFAR Eff} \end{tabular}}
                & \multirow{1}{*}{\begin{tabular}[c]{@{}l@{}} - \end{tabular}}
                & \multirow{1}{*}{\begin{tabular}[c]{@{}l@{}} - \end{tabular}}
                & \multirow{1}{*}{\begin{tabular}[c]{@{}l@{}} AIMC: \textbf{14.4 Top/s/W} \end{tabular}}
                & \multirow{1}{*}{\begin{tabular}[c]{@{}l@{}} - \end{tabular}}
                & \multirow{1}{*}{\begin{tabular}[c]{@{}l@{}} 6.38 Top/s/W (RBE mixed) \end{tabular}} \\
          \multirow{1}{*}{\begin{tabular}[c]{@{}l@{}} \textit{ResNet-20/CIFAR Lat} \tnote{g} \end{tabular}}
                & \multirow{1}{*}{\begin{tabular}[c]{@{}l@{}} - \end{tabular}}
                & \multirow{1}{*}{\begin{tabular}[c]{@{}l@{}} - \end{tabular}}
                & \multirow{1}{*}{\begin{tabular}[c]{@{}l@{}} 1.26ms \end{tabular}}
                & \multirow{1}{*}{\begin{tabular}[c]{@{}l@{}} - \end{tabular}}
                & \multirow{1}{*}{\begin{tabular}[c]{@{}l@{}} \textbf{1.05ms} \end{tabular}} \\
          \midrule
          \multirow{1}{*}{\begin{tabular}[c]{@{}l@{}} \textit{ResNet-18/ImageNet Eff} \end{tabular}}
                & \multirow{1}{*}{\begin{tabular}[c]{@{}l@{}} - \end{tabular}}
                & \multirow{1}{*}{\begin{tabular}[c]{@{}l@{}} - \end{tabular}}
                & \multirow{1}{*}{\begin{tabular}[c]{@{}l@{}} \textbf{19 Top/s/W} \end{tabular}}
                & \multirow{1}{*}{\begin{tabular}[c]{@{}l@{}} 12.1 Top/s/W \tnote{f} \end{tabular}}
                & \multirow{1}{*}{\begin{tabular}[c]{@{}l@{}} 5.83 Top/s/W (RBE 4\x{}4b) \end{tabular}} \\
          \multirow{1}{*}{\begin{tabular}[c]{@{}l@{}} \textit{ResNet-18/ImageNet Lat} \tnote{g}  \end{tabular}}
                & \multirow{1}{*}{\begin{tabular}[c]{@{}l@{}} - \end{tabular}}
                & \multirow{1}{*}{\begin{tabular}[c]{@{}l@{}} - \end{tabular}}
                & \multirow{1}{*}{\begin{tabular}[c]{@{}l@{}} \textbf{6.15ms} \end{tabular}}
                & \multirow{1}{*}{\begin{tabular}[c]{@{}l@{}} 24.8ms \end{tabular}}
                & \multirow{1}{*}{\begin{tabular}[c]{@{}l@{}} 48ms \end{tabular}} \\
          \bottomrule
        \end{tabular}
        \begin{tablenotes}[para, flushleft]
          \item [a] \textsc{cluster} area in brackets
          \item [b] architecture with 8\x{}RISC-V cores sharing 4\x{}FPUs
          \item [c] architecture with 1\x{}RISC-V core w/o FPUs
          \item [d] architecture with 16\x{}RISC-V cores sharing 8\x{} FPUs
          \item [f] zero-skipping
          \item [g] at best efficiency operating point.
        \end{tablenotes}
      \end{threeparttable}
      \label{tab:marsellus_rel_work}
  \end{adjustbox}
\end{table*}

The aggressive voltage scaling and body biasing capabilities of \marsellus{} enable further substantial energy savings by either lowering the supply voltage to \SI{0.65}{\volt} and activating \gls{abb} (down to $\sim$\SI{21}{\micro\joule}), or exploiting aggressive voltage scaling to \SI{0.5}{\volt} without \gls{abb} (down to \SI{12}{\micro\joule}).
The former technique has the advantage of inducing no performance penalty, whereas aggressive voltage scaling yields better energy efficiency but with 4$\times$ higher execution time.
Fig.~\ref{fig:fig12_resnet20_detail} details the layer-level performance and efficiency for this network in four operating points/precision configurations, while Fig.~\ref{fig:fig14_rebuttal} shows the detailed latency of off-chip (L3/L2), on-chip (L2/L1) and execution (RBE compute + tiling overheads).
Depending on the arithmetic intensity of each layer (number of operations performed per byte of data transferred), DMA transfers can be fully overlapped with execution (green labels) or bound by on-chip L2-L1 traffic (blue labels) or external L3-L2 traffic (red labels). In the latter two cases, the overhead paid is simply the difference between the DMA transfer time and the execution time.
Overall, architectural  heterogeneity and dynamic \gls{abb} enable opportunistically exploiting different key techniques for efficiency depending on the application needs.

\section{State-of-the-Art Comparison}
\label{sec:soa}





\subsection{ABB methods in the State-of-the-Art}
Table~\ref{tab:abb} compares the \gls{abb} strategy employed in the \marsellus{} prototype with other works in the \gls{soa}.
Most techniques achieve power savings in the order of 20--30\%; however, many of these works~\cite{moursy35021mm2PVTAware2021,akgulPowerManagementDVFS2014,quelen5mW0067mm2Automatic2018} target very simple prototypes where \gls{abb} is employed to regulate a simple digital core. 
Rossi~et~al.~\cite{rossi60GOPS8V2016} focus on a more complex architecture comprising a 4-core OpenRISC cluster and \SI{64}{\kibi\byte} of memory (which can be seen as a much simpler iteration of \marsellus{}'s \textsc{cluster}).
Finally, SleepRunner~\cite{bolSleepRunner28nmFDSOI2021} is a complete \gls{mcu} with a Cortex-M0 core, focusing on ultra-low-voltage and power execution.
Among these, \marsellus{} stands out as is it is arguably the most complex system on which \gls{abb} is applied.

Moreover, all comparable \gls{abb} methods, except for SleepRunner, apply their techniques due to offline analysis or after empirical trial-and-error.
SleepRunner and \marsellus{} are unique in featuring on-chip techniques to automatically tune \gls{abb} according to runtime requirements.
In the case of SleepRunner, this happens with a technique caled Unified Frequency/Biasing Regulation (UFBR), which is based on a centralized ring oscillator within the regulator to track the actual operating frequency with respect to the set point, therefore compensating \gls{pvt} variations.
\marsellus{}'s technique is unique as, thanks to \glspl{ocm}, it can track the operating frequency in an application-specific way.

\subsection{SoCs targeted at AI-IoT}
The \marsellus{} \gls{soc} is positioned in the \gls{soa} as part of an industry trend toward architectural heterogeneity to cope with \gls{ai}-\gls{iot} workloads. Commercial devices following this pattern include, for example, Analog Devices MAX78000 (which combines a Cortex-M4 \gls{mcu} with a custom \gls{npu})\footnote{\url{https://www.analog.com/en/products/max78000.html}}, the Syntiant NDP120 (Cortex-M0 \gls{mcu} + Syntiant Core 2 \gls{npu})\footnote{\url{https://www.syntiant.com/ndp120}}, the upcoming STMicroelectronics STM32N6 (Cortex-M class \gls{mcu} + Neural-Art \gls{npu})\footnote{\url{https://blog.st.com/stm32n6/}}, the Alif Semiconductor Ensemble family (Cortex-M55 + Ethos-U55 \gls{npu}\footnote{https://alifsemi.com/ensemble/}), and the GreenWaves Technologies GAP9 (RISC-V \gls{mcu} + 9\x{} RISC-V \gls{dsp} cluster + NE16 Neural Engine)\footnote{https://greenwaves-technologies.com/gap9\_processor/}. Except for the latter, all of these devices dedicate most silicon area to \gls{dnn} acceleration, leaving non-\gls{ml} workloads to execution on the microcontroller.
Both \marsellus{} and the commercial GAP9 \gls{soc} take a different approach, employing efficient (but inflexible) hardware acceleration engines (\gls{rbe} and NE16, respectively) and more flexible (but less efficient) parallel acceleration engines (the RISC-V clusters). The reason for this choice is that, by Amdahl’s law, even if a small percentage of the original non-accelerated workload is non-\gls{ml}, with hardware acceleration this can become a major performance bottleneck.

We focus quantitative comparisons on other non-commercial research prototypes.
Table~\ref{tab:marsellus_rel_work} compares \marsellus{} with four recently
presented \gls{soa} \glspl{soc} targeted at emerging AI-IoT applications: Vega~\cite{rossiVegaTenCoreSoC2022}, an AI-IoT 10-core \gls{soc} (precursor to the GAP9 \gls{soc}) with legacy \gls{cnn} acceleration capabilities; SamurAI~\cite{miro-panadesSamurAIVersatileIoT2022}, a single-core RISC-V microcontroller with a powerful embedded \gls{dnn} accelerator; DIANA~\cite{houshmandDIANAEndtoEndHybrid2023}, a \gls{soc} with hybrid \gls{aimc} and digital hardware acceleration capabilities; and QNAP~\cite{mo12TOPSQuantized2022}, an 8-bit \gls{dnn}-dedicated \gls{asic} with zero-skipping capabilities.
All \glspl{soc} are fabricated in \SI{22}{\nano\meter} or \SI{28}{\nano\meter} technology nodes, reducing the impact of pure technology scaling on the comparison.

The contribution of \marsellus{} stands out in several dimensions.
First, \marsellus{} provides significantly larger software performance while keeping within a power range comparable with that of the other \glspl{soc}.
This is due to the combined effects of the 16-core architecture, aggressive 2-bit quantization with \texttt{Xpulpnn}, and \gls{abb}, which are features available only in \marsellus{}.
Second, while most of the \glspl{soc} in the table employ heterogeneous architectures, \marsellus{} is more completely pushing in the direction of combining high-performance software with flexible acceleration, thanks to the fine-grain precision support in \gls{rbe}.
Combined, these effects yield \gls{soa}-leading results in terms of performance, area efficiency, and energy efficiency for software (INT and FP); hardware-accelerated execution is leading or equivalent in terms of both performance and energy efficiency compared to the other digital hardware-accelerated \glspl{soc}, and it is second in terms of area efficiency after QNAP, which is not surprising as the latter is a dedicated \gls{asic}.
DIANA's \gls{aimc} accelerator provides (as could be expected) peak performance and efficiency metrics that are $\sim$100$\times$ better than digital accelerators; however, compared with digital accelerators, \gls{aimc} units are significantly harder to utilize efficiently, as also noted by the architects of DIANA~\cite{vandelmHTVMEfficientNeural2023}: this means that such metrics are difficult to compare fairly with those of digital accelerators.

To deepen the perspective of the comparison between the proposed architectures, Table~\ref{tab:marsellus_rel_work} also includes performance and efficiency achieved by the various \glspl{soc} on two common benchmarks for \gls{cnn} accelerators, namely, ResNet-20/CIFAR (using the scheme discussed in Section~\ref{sec:dnn}) and ResNet-18/ImageNet (targeting a full HAWQ-quantized 4-bit network, achieving 68.5\% on ImageNet\footnote{https://github.com/Zhen-Dong/HAWQ}).
Interestingly, these results confirm that the difference in efficiency between digital and \gls{aimc} accelerators is significantly reduced, from the theoretical $\sim$50--100\x{} to a 2.3--5\x{} advantage in practice.
Similar considerations apply to peak performance.
This is due to limited utilization of the \gls{aimc} module in practical layers, as well as overheads from digital periphery.
In terms of performance, in the case of ResNet-20/CIFAR inference, \marsellus{} can actually deliver 20\% faster runtime due to the same effects.
On ResNet-18/ImageNet the results of DIANA are significantly better (likely due to higher utilization of the \gls{aimc} array compared to the ResNet-20/CIFAR case); QNAP delivers the second-best result.
Overall, our work is very competitive even when considering architectures with aggressive approximation from \gls{aimc} and micro-optimizations such as zero-skipping, which are not used in \marsellus{} -- while at the same time offering more flexibility as non-hardware-accelerated tasks can be executed in software at \gls{soa}-leading performance and efficiency on the ISA-enhanced RISC-V cores.





\section{Conclusion}
\label{sec:conclusion}

This work presented \marsellus{}, an advanced AI-IoT \gls{soc} fabricated in \SI{22}{\nano\meter} FDX technology combining a heterogeneous architecture with a cluster of 16 RISC-V cores with advanced \gls{dsp} and \gls{ai} \gls{isa} extensions, fully integrated with a flexible-precision partially bit-serial \gls{dnn} accelerator.
The \marsellus{} \gls{soc} can be aggressively voltage- and frequency-scaled to improve energy efficiency, and the dynamic \gls{abb} mechanism introduced in this prototype enables fine-grained tuning of performance, and efficiency optimization even without scaling frequency.
Fig.~\ref{fig:fig13_conclusion} summarizes all the efficiency optimization techniques discussed in this work in terms of energy per elementary operation.
The combination of architecture improvements, data quantization, and voltage scaling with body biasing or frequency scaling yield a plethora of different options for energy vs flexibility/accuracy trade-offs on the same \gls{soc}.
\marsellus{} responds to the demand for architectural flexibility and capability to adapt the same computing fabric to diverse tasks, as required by current and future AI-IoT applications.

\begin{figure}[tb]
    \centering\includegraphics[width=0.99\columnwidth]{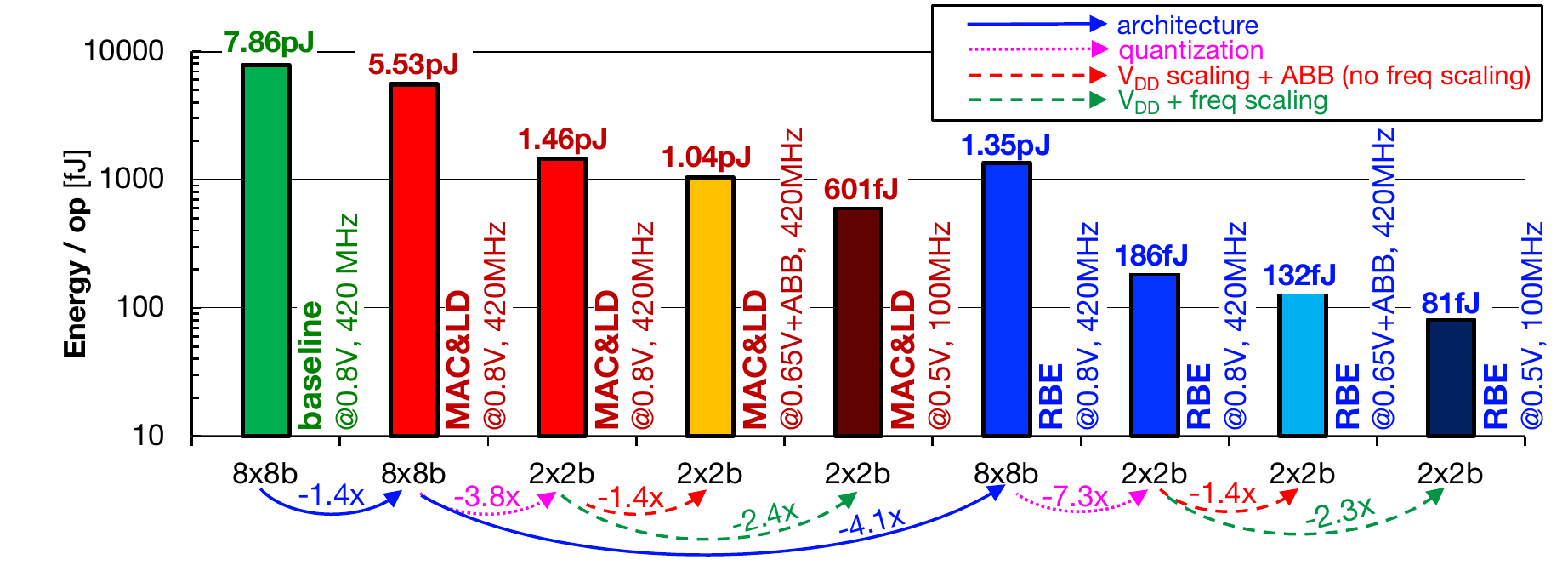}
    \caption{Summary of energy efficiency optimization techniques available in \marsellus{}.}
    \label{fig:fig13_conclusion}
\end{figure}

The baseline \gls{rtl} code of the \textsc{cluster}\footnote{\url{https://github.com/pulp-platform/pulp}} and of the \gls{rbe} accelerator\footnote{\url{https://github.com/pulp-platform/rbe}} are released as open-source under a liberal license to foster future research in the area of AI-IoT computing devices.



\section*{Acknowledgment}
We thank Dolphin Design, France, for collaborating with the implementation and fabrication of the \marsellus{} silicon prototype and all measurements.



\bibliographystyle{IEEEtran}

%



%

\begin{IEEEbiography}[{\includegraphics[width=0.95in,height=1.25in,clip,keepaspectratio]{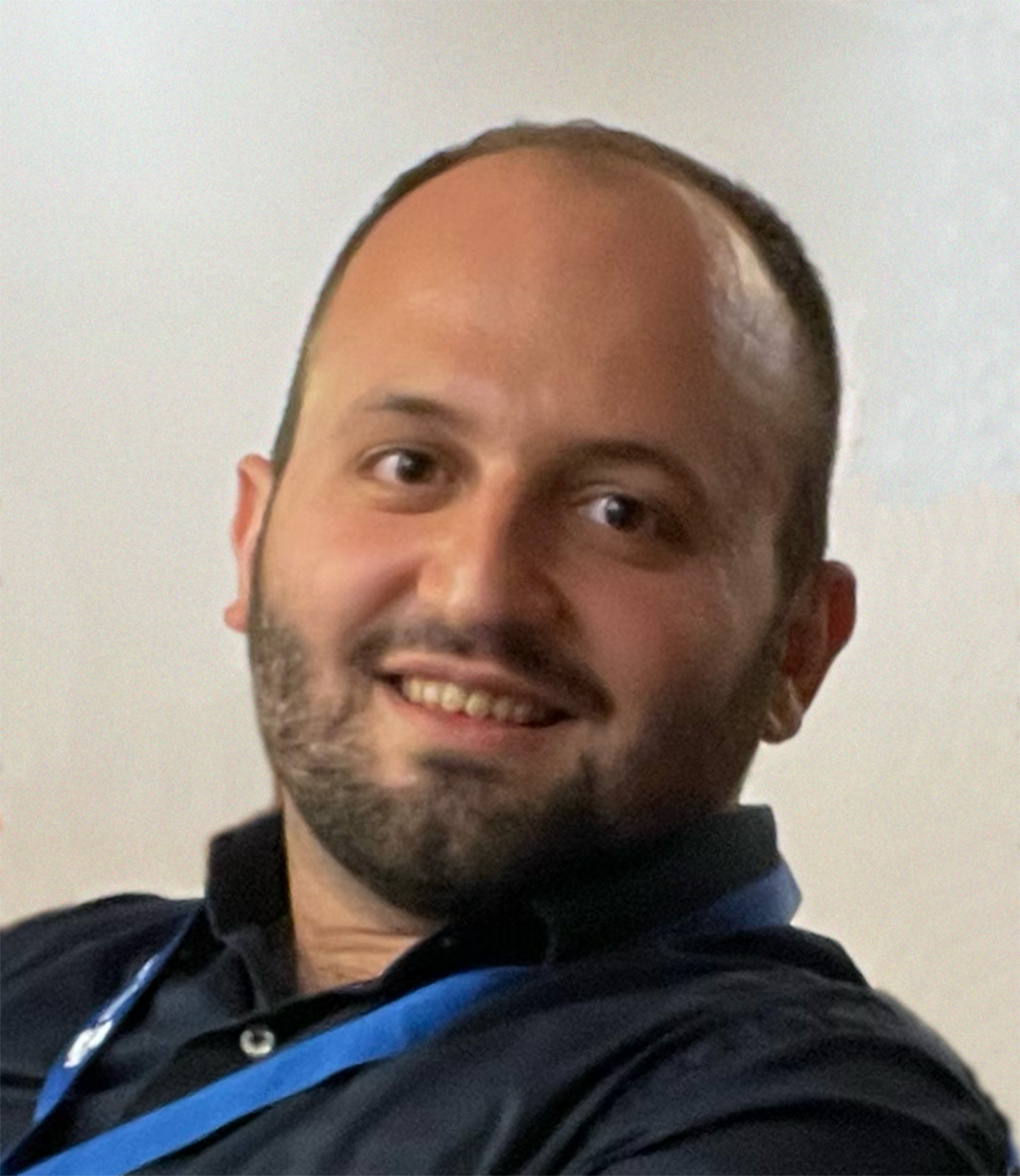}}]{Francesco Conti} (Member, IEEE) received the Ph.D. degree in electronic engineering from the University of Bologna, Italy, in 2016.
He is currently a Tenure-Track Assistant Professor with the DEI Department, University of Bologna. From 2016 to 2020, he held a research grant with the University of Bologna and a Post-Doctoral Researcher with ETH Zürich.
His research is centered on hardware acceleration in ultra-low power and highly energy efficient platforms, with a particular focus on System-on-Chips for Artificial Intelligence applications.
His research work has resulted in more than 70 publications in international conferences and journals and was awarded several times, including the 2020 IEEE \textsc{Transactions on Circuits and Systems I: Regular Papers} Darlington Best Paper Award.
\end{IEEEbiography}

\begin{IEEEbiography}[{\includegraphics[width=1.05in,height=1.25in,clip,keepaspectratio]{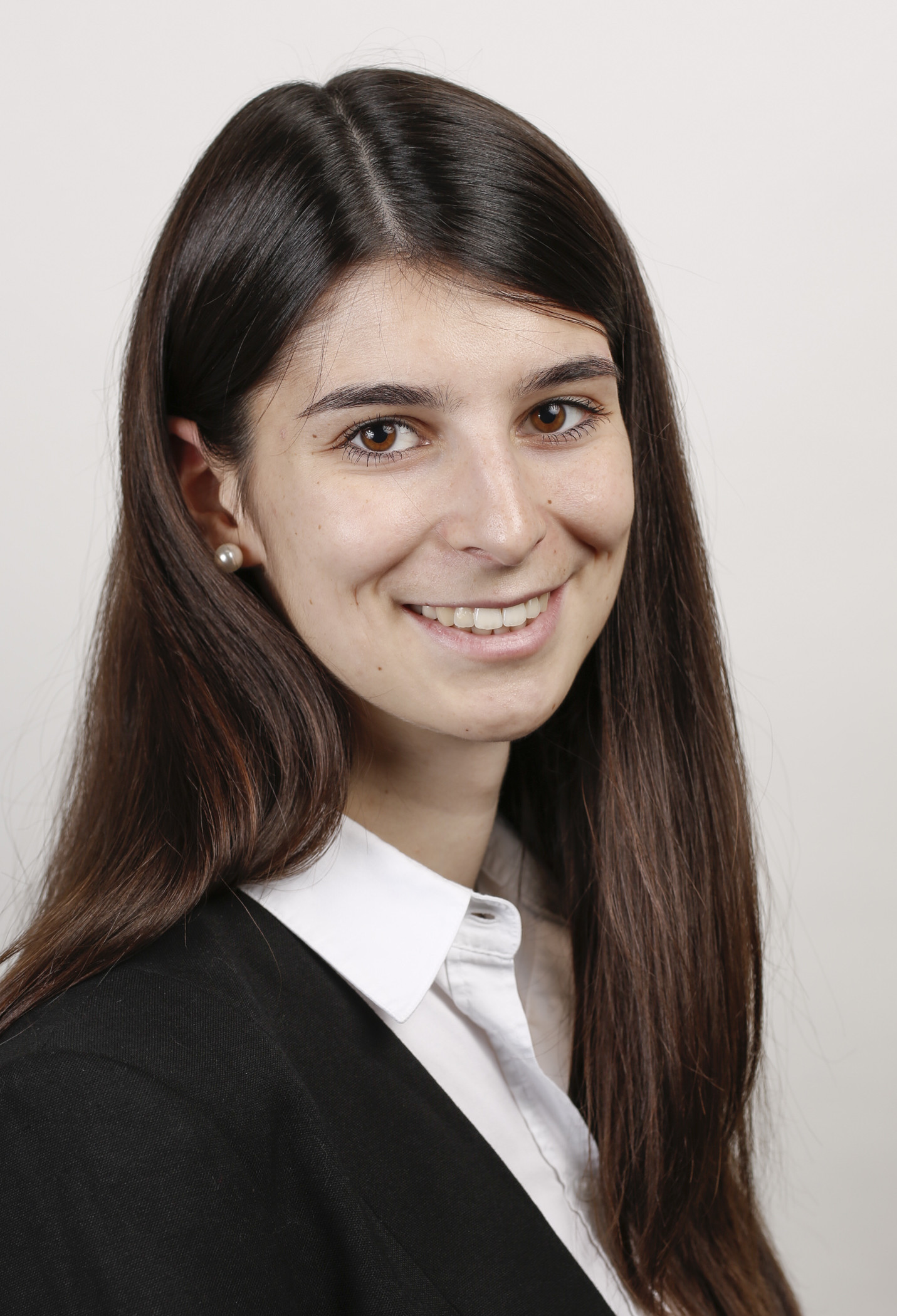}}]{Gianna Paulin} received her BSc and MSc in ``Electrical Engineering and Information Technology'' from the Swiss Federal Institute of Technology Z\"urich (ETHZ), Switzerland, in 2017 and 2019, respectively. In 2019 she joined the Integrated Systems Laboratory of ETH Z\"urich as a PhD candidate. Her research interests include computer architecture and hardware acceleration of deep learning applications targeting both, low power embedded systems and high-performance computing systems.
\end{IEEEbiography}

\begin{IEEEbiography}[{\includegraphics[width=1.05in,height=1.25in,clip,keepaspectratio]{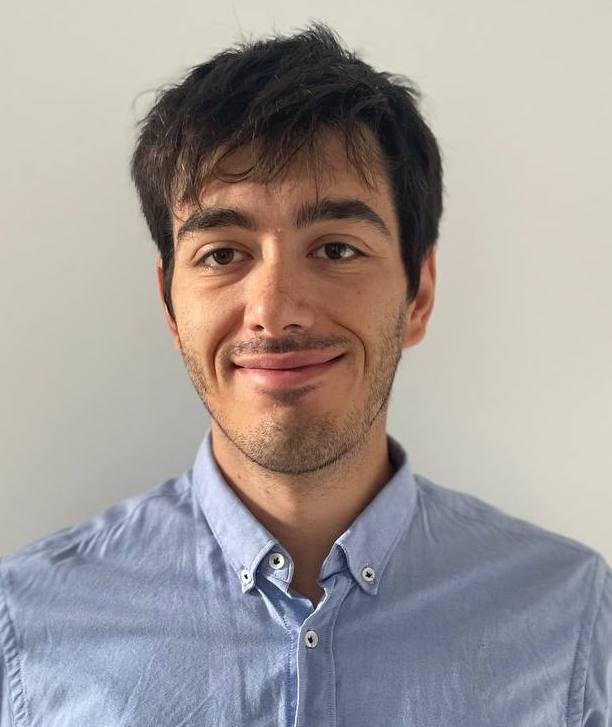}}]{Angelo Garofalo} (Member, IEEE) received the B.Sc., M.Sc., and Ph.D. degrees in electronic engineering from the University of Bologna, Italy, in 2016 and 2018, and 2021, respectively.
He is currently an Assistant Professor with the Department of Electrical, Electronic and Information Engineering (DEI). His main research topic is hardware–software design of ultra-low-power multiprocessor systems on chip for edge AI. His research interests include quantized neural networks, hardware efficient machine learning, in-memory computing, heterogeneous architectures, and fully programmable embedded architectures.
\end{IEEEbiography}

\begin{IEEEbiography}[{\includegraphics[width=1.05in,height=1.25in,clip,keepaspectratio]{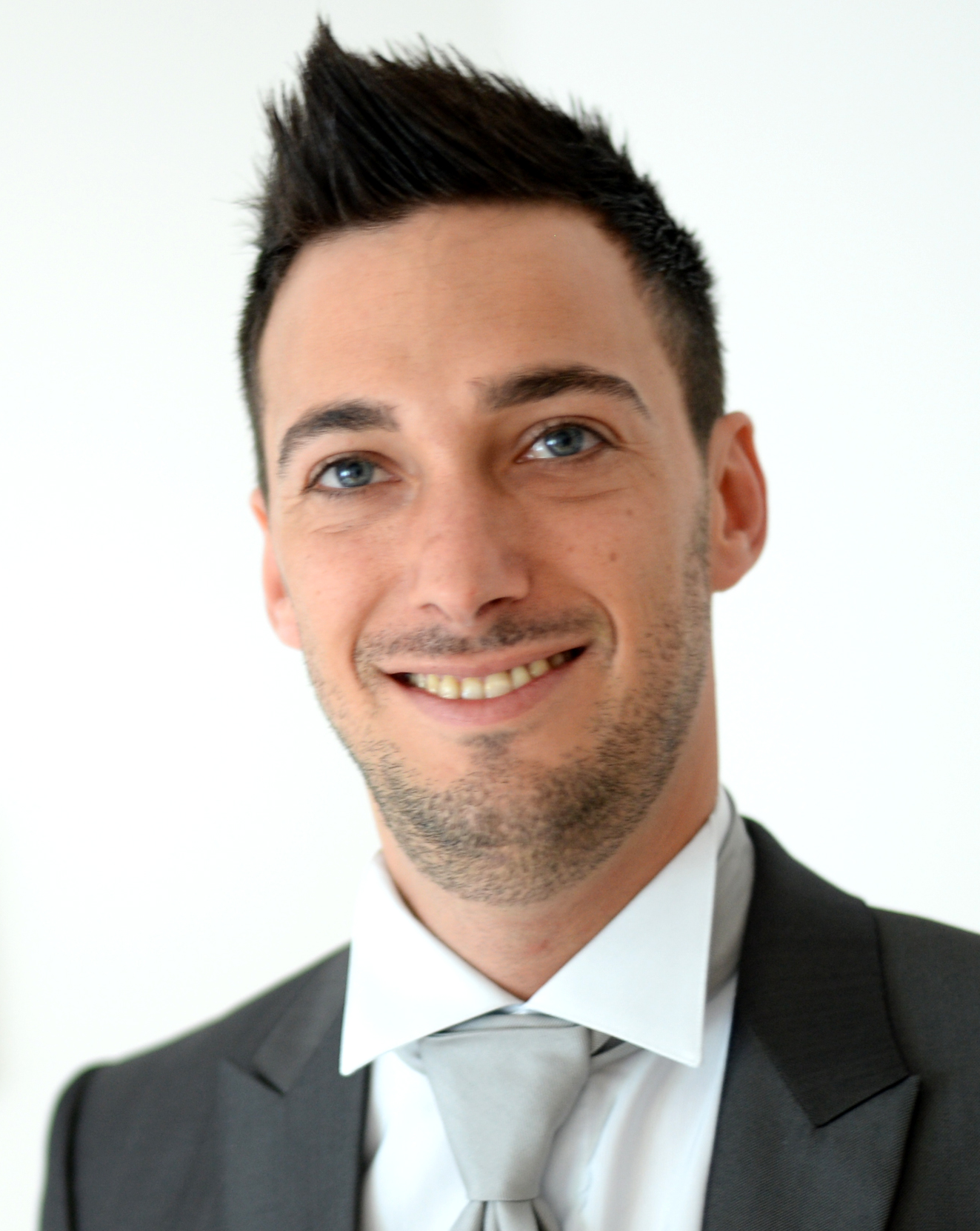}}]{Davide Rossi} (Member, IEEE) received the Ph.D. degree from the University of Bologna, Bologna, Italy, in 2012. He has been a Post-Doctoral Researcher with the Department of Electrical, Electronic and Information Engineering ``Guglielmo Marconi,'' University of Bologna, since 2015, where he is currently an Associate Professor. His research interests focus on energy-efficient digital architectures. In this field, he has published more than 100 papers in international peer-reviewed conferences and journals. He was a recipient of the Donald O. Pederson Best Paper Award 2018, the 2020 IEEE \textsc{Transactions on Circuits and Systems} Darlington Best Paper Award, and the 2020 IEEE \textsc{Transactions on Very Large Scale Integration (VLSI) Systems} Prize Paper Award.
\end{IEEEbiography}


\begin{IEEEbiography}[{\includegraphics[width=1.05in,height=1.25in,clip,keepaspectratio]{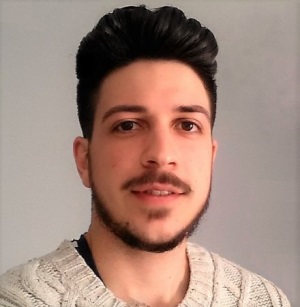}}]{Alfio Di Mauro} (Member, IEEE) received the M.Sc. degree in electronic engineering from the Electronics and Telecommunications Department (DET), Politecnico di Torino, in 2016, and the Ph.D. degree with the Integrated System Laboratory (IIS), Swiss Federal Institute of Technology, Zürich, in 2021.
His research focuses on the design of digital ultra-low power (ULP) system-on-chip (SoC) for event-driven edge computing.
\end{IEEEbiography}

\begin{IEEEbiography}[{\includegraphics[width=1.05in,height=1.25in,clip,keepaspectratio]{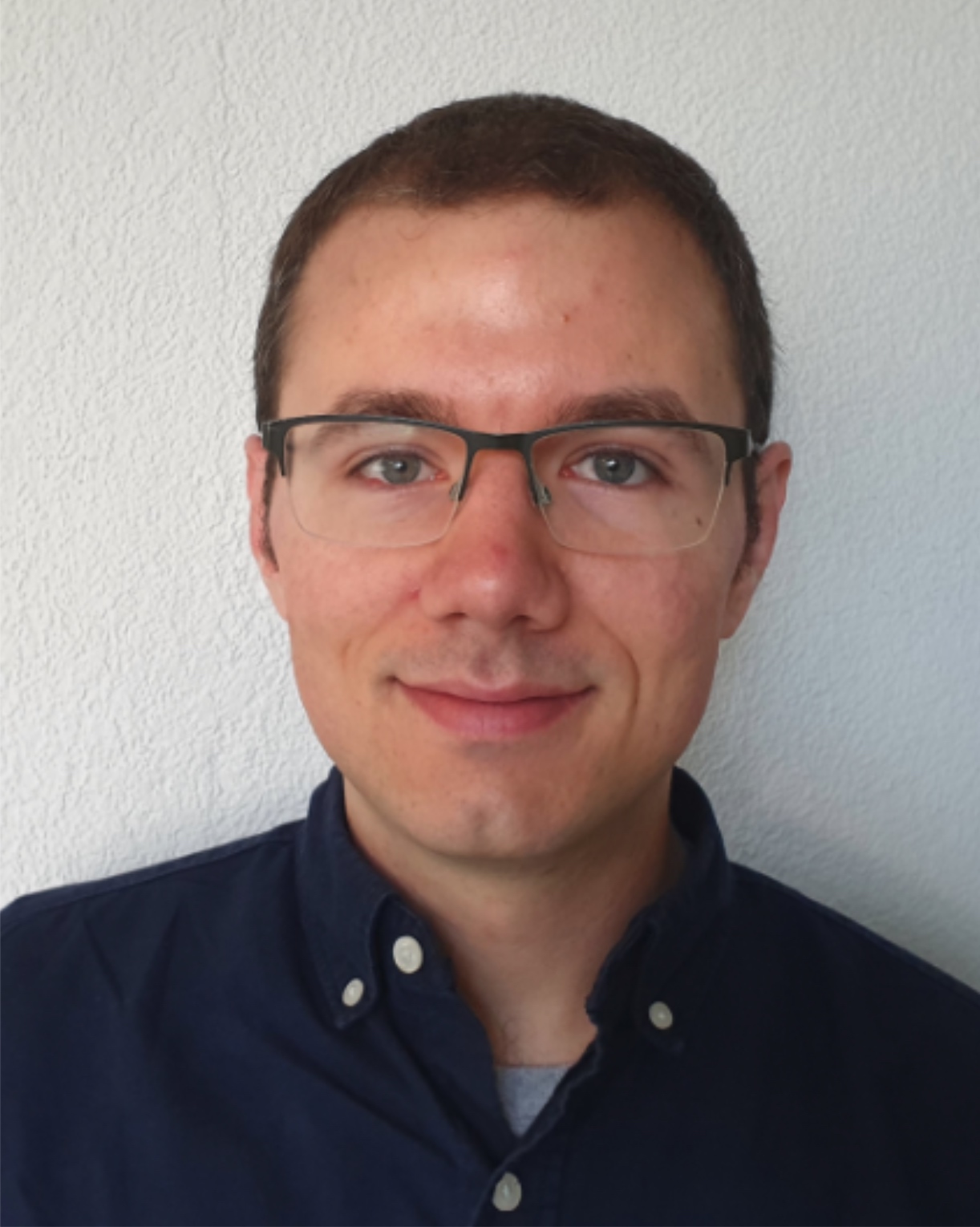}}]{Georg Rutishauser} (Graduate Student Member, IEEE) received the B.Sc. and M.Sc. degrees in electrical engineering and information technology from ETH Z\"urich, Z\"urich, Switzerland, in 2015 and 2018, respectively, where he is currently pursuing the Ph.D. degree with the Integrated Systems Laboratory.
His research interests include algorithms and hardware for reduced-precision deep learning and their application in computer vision and embedded systems.
\end{IEEEbiography}

\begin{IEEEbiography}[{\includegraphics[width=1.05in,height=1.25in,clip,keepaspectratio]{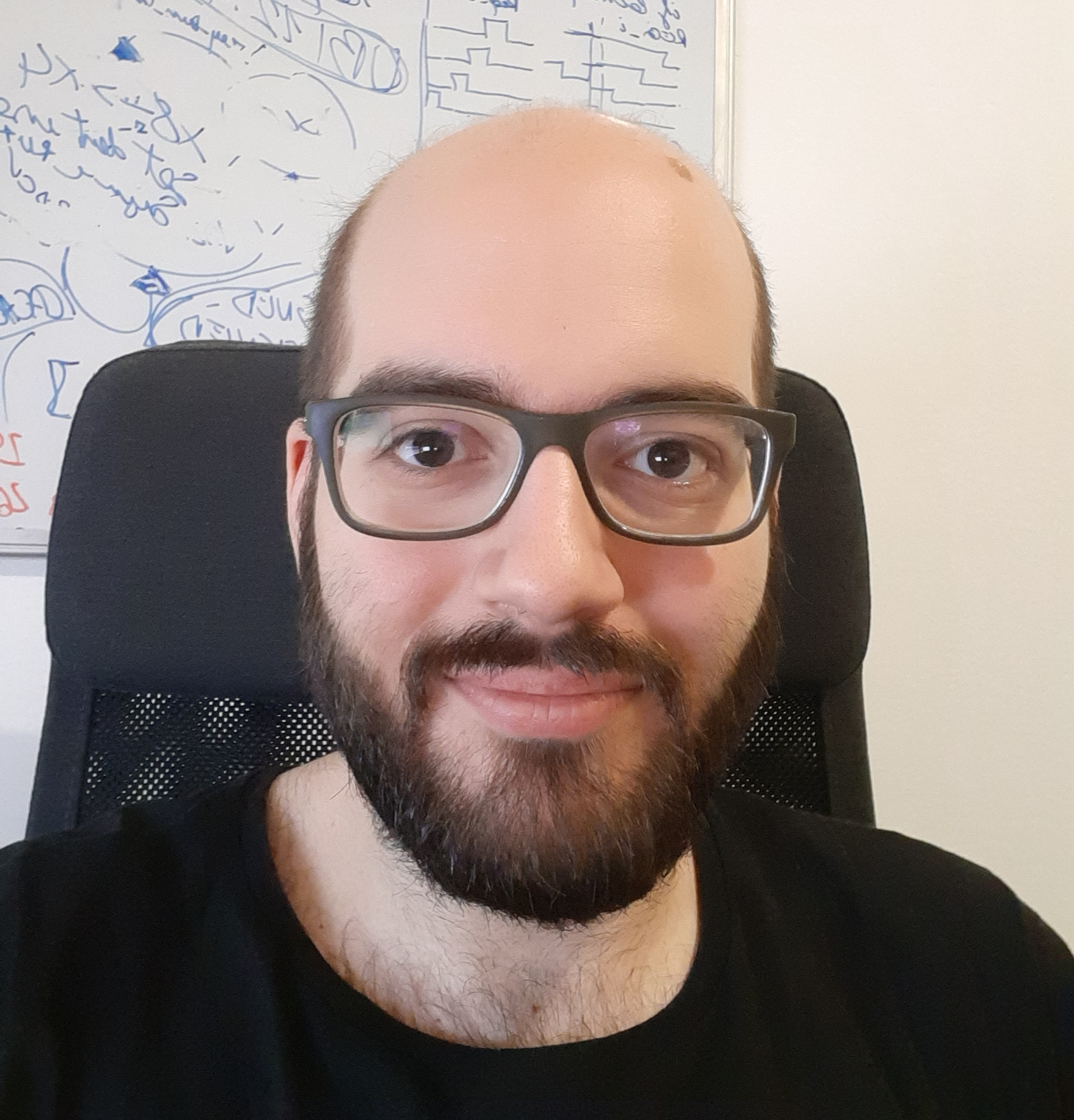}}]{Gianmarco Ottavi} received the M.Sc. degree in 2019. He is currently pursuing the Ph.D. degree in electronics engineering with the University of Bologna, Italy. He was a Research Fellow with the Department of Electrical, Electronic and Information Engineering (DEI), Bologna, for two years.
His research is focused on hardware design for efficient inference in low-power systems, where he developed specialized ISA extensions for RISC-V and system-level implementation of in-memory computing accelerators.
\end{IEEEbiography}

\begin{IEEEbiography}[{\includegraphics[width=1.05in,height=1.25in,clip,keepaspectratio]{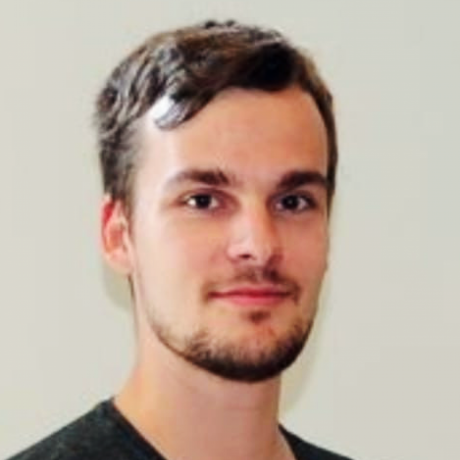}}]{Manuel Eggimann} (Member, IEEE) received the M.Sc. degree in electrical engineering and information technology from ETH Z\"urich, Z\"urich, Switzerland, in 2018, where he is currently pursuing the Ph.D. degree with the ETH Z\"urich Integrated Systems Laboratory.
His research interests include low-power hardware design, edge computing, and very-large-scale integration (VLSI).
Mr. Eggimann was a recipient of the Best Paper Award at the 2019 IEEE 8th International Workshop on Advances in Sensors and Interfaces.
\end{IEEEbiography}

\begin{IEEEbiography}[{\includegraphics[width=1.05in,height=1.25in,clip,keepaspectratio]{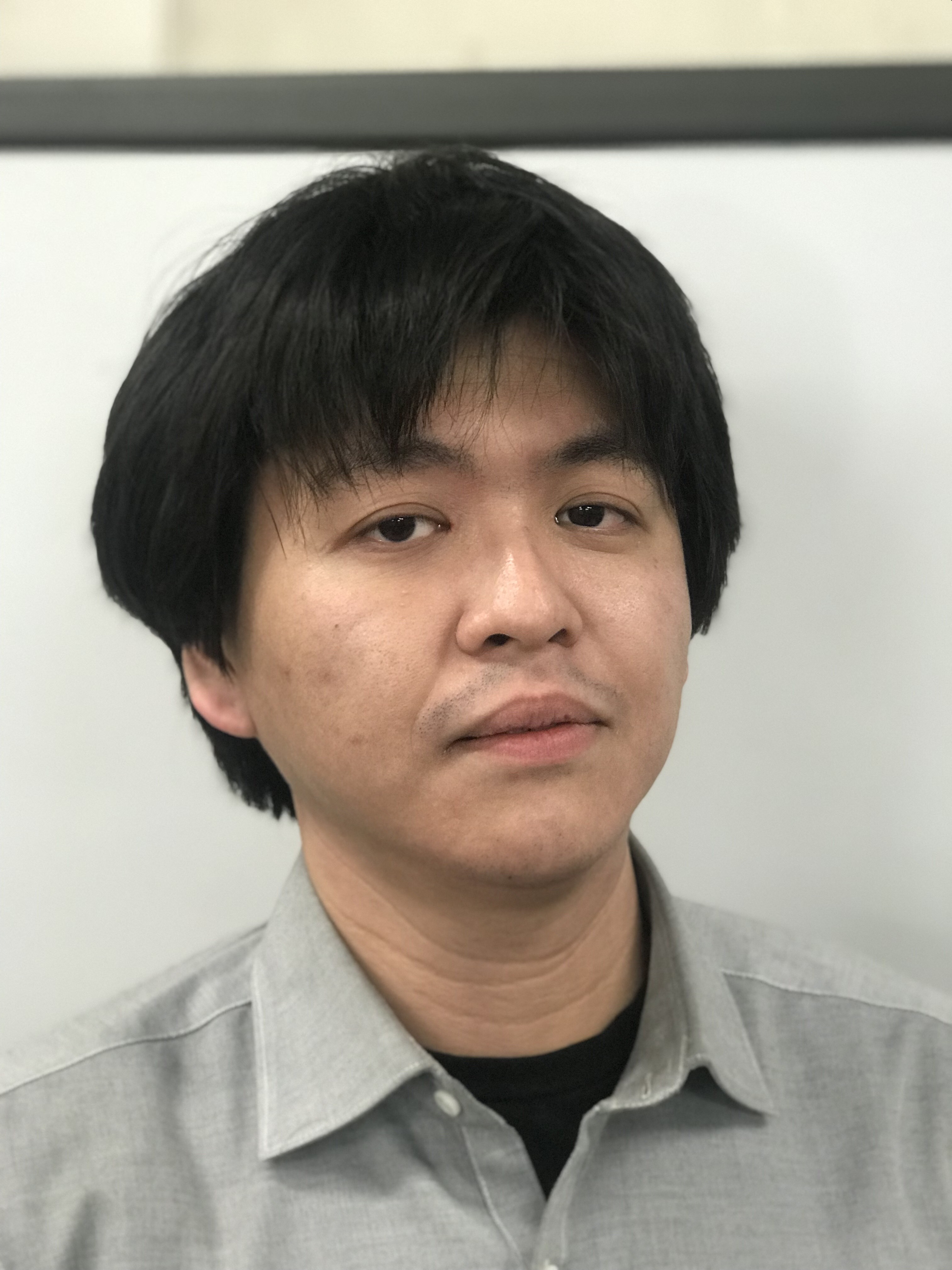}}]{Hayate Okuhara} (Member, IEEE) received the Ph.D. degree from Keio University, Kanagawa, Japan, in 2018. He has been a Postdoctoral Researcher with the Department of Electrical, Electronic and Information Engineering ``Guglielmo Marconi,'' University of Bologna, Bologna, Italy till 2021 and is currently with the Department of Electrical and Computer Engineering, National University of Singapore, Singapore. His research interest includes low-power VLSI system design.
\end{IEEEbiography}

\begin{IEEEbiography}[{\includegraphics[width=1.05in,height=1.25in,clip,keepaspectratio]{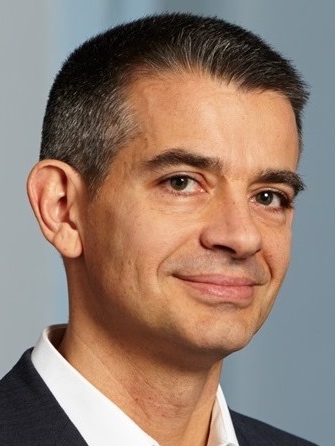}}]{Luca Benini} (Fellow, IEEE) holds the chair of digital Circuits and systems at ETHZ and is Full Professor at the Universit\`a di Bologna. He received a PhD from Stanford University.  Dr. Benini’s research interests are in energy-efficient parallel computing systems, smart sensing micro-systems and machine learning hardware. He is a Fellow of the ACM and a member of the Academia Europaea. He is the recipient of the 2016 IEEE CAS Mac Van Valkenburg award, the 2020 EDAA achievement Award, the 2020 ACM/IEEE A. Richard Newton Award and the 2023 IEEE CS E.J. McCluskey Award.
\end{IEEEbiography}




\end{document}